\documentclass[twocolumn]{aastex63}

\usepackage{graphicx}	% Including figure files
\usepackage{amsmath}	% Advanced maths commands
\usepackage{amssymb}	% Extra maths symbols
\usepackage{units}

\received{May 18, 2020}
\revised{July 3, 2020}
\accepted{July 7, 2020}
\published{August 28, 2020}
\submitjournal{ApJ}

\graphicspath{{./}{figures/}}
\begin{document}

\title{Gas-driven inspiral of binaries in thin accretion disks}

\author[0000-0002-3820-2404]{Christopher Tiede}
\affiliation{Center for Cosmology and Particle Physics, Physics Department, New York University, New York, NY 10003, USA}
\email{cwt271@nyu.edu}

\author{Jonathan Zrake}
\affiliation{Department of Physics and Astronomy, Clemson University, SC 29634, USA}

\author[0000-0002-0106-9013]{Andrew MacFadyen}
\affiliation{Center for Cosmology and Particle Physics, Physics Department, New York University, New York, NY 10003, USA}

\author{Zoltan Haiman}
\affiliation{Department of Astronomy, Columbia University, New York, NY 10027, USA}



\begin{abstract}

Numerical studies of gas accretion onto supermassive black hole binaries (SMBHBs) have generally been limited to conditions where the circumbinary disk (CBD) is 10--100 times thicker than expected for disks in active galactic nuclei (AGN). This discrepancy arises from technical limitations, and also from publication bias toward replicating fiducial numerical models. Here we present the first systematic study of how the binary's orbital evolution varies with disk scale height. We report three key results: (1) Binary orbital evolution switches from outspiralling for warm disks (aspect ratio $h/r \sim 0.1$), to inspiralling for more realistic cooler, thinner disks at a critical value of $h/r \sim 0.04$, corresponding to orbital Mach number $\mathcal{M}_{\rm crit}\approx25$. (2) The net torque on the binary arises from a competition between positive torque from gas orbiting close to the black holes, and negative torque from the inner edge of the CBD, which is denser for thinner disks. This leads to increasingly negative net torques on the binary for increasingly thin disks. (3) The accretion rate is modestly suppressed with increasing Mach number. We discuss how our results may influence modeling of the nano-Hz gravitational wave background, as well as estimates of the LISA merger event rate.

\end{abstract}



% ======================================================================
\section{Introduction}

There is strong evidence that most galaxies host a super-massive black hole (SMBH) in their center \citep[e.g][]{Dressler1988, Kormendy1995, Ferrarese2005}. Furthermore, in the prevailing paradigm of hierarchical structure  formation in the universe, larger, more complex architectures grow from the mergers and interactions of smaller ones. It follows that we ought to expect super-massive black hole binaries (SMBHBs) to be a natural consequence of galaxy mergers \citep{KormendyHo2013, Woods+2019}. Following one of these galaxy mergers, dynamical friction from gravitational interactions with surrounding dark matter and stars is expected to drive the SMBHs into bound pairs and lead to compact binaries with orbital separations of order $\sim 1\text{pc}$ in the new galactic nucleus \citep[e.g.][]{Begel:Blan:Rees:1980, Roos1981, 2001ApJ...563...34M, Merritt2005}. These post-merger galaxies are also expected to have ample gas in their nuclei \citep{BarnesHernquist1996} and thus, a central SMBHB would likely be surrounded by and interact with this fluid~\citep{Springel2005}. The fluid is expected to radiate efficiently, and to collapse along its axis of rotation into a thin co-rotating accretion disk \citep{SS1973}. This circumbinary accretion disk (CBD), and its interaction with the SMBHB, have been the subject of intense theoretical and numerical studies. In certain situations, the CBD has been found to facilitate binary's orbital decay down to separations at which gravitational radiation becomes the dominant mechanism, driving the SMBHB to merger \citep[e.g.][]{Armitage2002, Escala2005, Dotti2007, Mayer+2007, 2008ApJ...672...83M, Dotti2009, HKM09, Fiacconi+2013, MML17, SouzaLima+2017, Yike17}.

One major uncertainty in this picture is whether and how rapidly the binary-disk interaction (or other mechanisms, such as scattering of stars entering the loss cone; e.g. \citealt{Vasiliev+2015} and references therein) can shrink the binary from $\sim \unit[1]{pc}$ down to $\sim \unit[10^{-2}]{pc}$ separation at which energy is lost to gravitational radiation fast enough for the binary to merge in a Hubble time --- commonly referred to as the ``final parsec problem'' \citep{fpp2003}. Therefore, a quantity of great interest in studying the binary-disk interaction is the net transfer of angular momentum between the binary and the disk, $\dot{L}$, and particularly the sign of the resulting evolution of the binary separation, $\dot{a}$.  Note that even if stellar scattering (or some other process) ``solves'' the final parsec problem, interaction with the circumbinary gas may still often be the dominant mechanism determining the binary's orbital evolution at small separations~\citep{HKM09, Kelley+2017a}. The gas-driven orbital evolution for compact SMBHBs is integral for estimating the merger rates of SMBHBs detectable by the space-based Laser Interferometer Space Antenna (LISA), as well as for modelling the gravitational wave background we expect to measure with Pulsar Timing Arrays~\citep{KocsisSesana2011, Kelley+2017b}. It also has direct implications for the interpretation of binary searches in large optical time-domain surveys~\citep{HKM09,Kelley+2019}, such as recently performed in the Catalina Real-Time Transient Survey~\citep[CRTS;][]{Graham+2015} and the Palomar Transient Factory \citep[PTF;][]{Charisi+2016}, and as expected in the forthcoming Legacy Survey of Space and Time (LSST) with the Vera C. Rubin Observatory.

Many of the early studies of the binary-disk interaction suggested that the disk removes angular momentum from the binary, appealing to a tidal-viscous interaction that distorts the nearby disk and deposits angular momentum, with the angular momentum then carried outward in the disk by enhanced viscous stresses \citep{SyerClarke1995, Gould2000, Armitage2002, Armitage2005, 2008ApJ...672...83M, HKM09, Kocsis+2012a, Kocsis+2012b, Rafikov2016}. A few early hydrodynamical simulations measured the torques due to the circumbinary disk, and found them to typically drive the binary inward, although the gas close to the BHs, in the central cavity surrounding the binary, was either poorly resolved~\citep{Cuadra2009, Roedig2012} or excised from the computational domain \citep{2008ApJ...672...83M, Shi+2012,MML17}.

Several works have recently begun studying the binary-disk interaction in more detail, by simulating binary-disk systems that place the binary in the simulation domain and better resolve the innermost region. In 2D simulations, \cite{Yike17} found that for physically motivated mass-removal rates in the sink particles representing the BHs, the gravitational torque exerted on the binary by the disk was negative, promoting inspiral. However, they noted that unlike in previous studies, the torque was dominated by the asymmetrically distributed gas near the edges of the ``mini-disks'' of the individual BHs, rather than by non-axisymmetric features in the circumbinary disk farther out. They also noted that for more rapid sink rates, these torques become positive. In similar 2D simulations, \cite{MML19} also found the gas near the individual BHs to dominate the torques, and obtained positive gravitational torques, supplying the binary with angular momentum and driving it apart.  In both 2D and 3D simulations, \cite{Moody19} similarly found positive net torques exerted on the binary and that these torques result in a positive $\dot{a}$, i.e. binary expansion. \cite{Moody19} additionally found that inclined disks yield positive torques and expanding binaries, concluding that viscous disks will deposit angular momentum into the binary and drive it apart ``in all cases''.

Both \cite{Moody19} and \cite{MML19} studied infinite disks (an inflow boundary condition at a suitably far distance from the inner-binary) for thousands of binary orbits (a few viscous times) in order to reach a steady-state.  Most recently, \cite{Munoz19}, examined the modifications to the steady-state results if one considered a finite disk instead.  They found that while the disk viscously expands and is gradually depleted by the accreting binary such that the system never reaches a steady-state, the angular momentum transferred to the binary per unit mass accreted, $\dot{L} / \dot{M}$, as well as the change in binary separation per unit mass accreted $\dot{a} / \dot{M}$ both settle to constant values (after an initial transient period). They also found that these values measured from a finite disk are in close agreement with those determined from an infinite disk in steady state \citep{MML19, Moody19}. Therefore, \cite{Munoz19} suggested that a true steady state is not required to calculate the binary torque and the evolution of binary separation.  Lastly, \cite{Munoz19} noted that the above results only depend very weakly on the value of the turbulent viscous parameter $\alpha$.

\begin{figure*}
    \centering
    \includegraphics[width=0.8\textwidth]{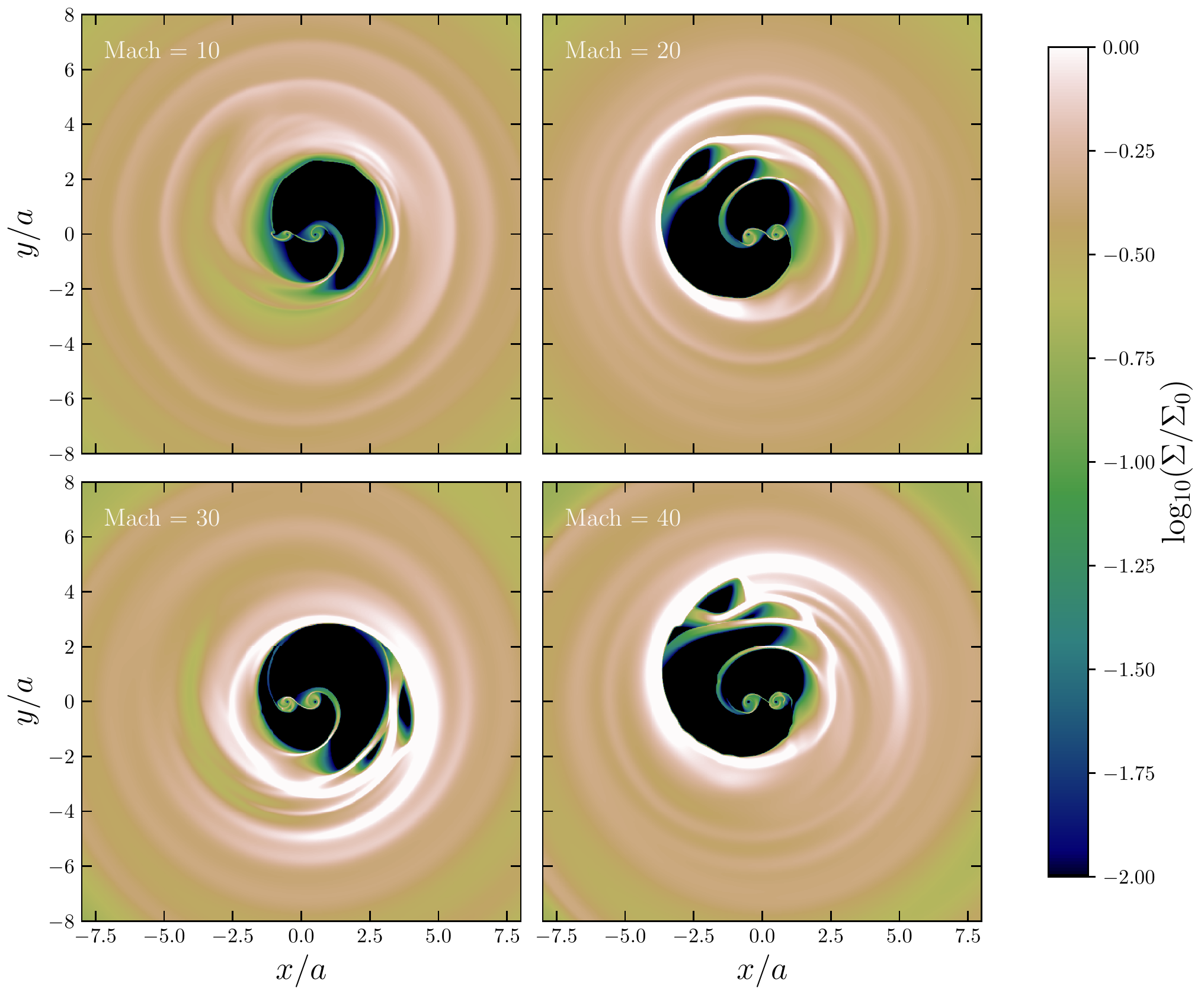}
    \protect\caption{Snapshots of the surface density distribution for four different values of the Mach number $\mathcal{M}$ after about $500$ orbits. 
    Qualitatively, with increasing $\mathcal{M}$ we see the development of slightly larger cavities as well as a marked growth in the pile-up of material at the cavity wall.}
    \label{fig:reliefs}
\end{figure*}

One major shortcoming of all of the aforementioned numerical studies,
however, is that they assume a disk aspect ratio of $h/r = 0.1$, which is one to two orders of magnitude thicker than current estimates of disk thickness in galactic centers. Mid-plane temperature measurements from accretion disks in AGN are markedly cooler and imply aspect ratios of order $(h/r) \sim 10^{-2} - 10^{-3}$ \citep[e.g.][]{krolikbook,Hubeny2001}.
The reasons for simulating unrealistically thick disks, though, are two-fold. First, for disks with $(h/r) \approx 10^{-3}$ (equivalent to a Mach number of $\mathcal{M} \approx 10^3$ for the azimuthal gas velocity) it is computationally difficult to 
resolve and follow high-contrast regions and to ensure that the density everywhere remains positive.
Second, preserving a standard thickness across studies is valuable for comparison within the literature. On the other hand, this may lead to inaccurate predictions, if important disk properties depend significantly on disk thickness.

The question of disk thickness was considered by \cite{Ragusa+2016}, who performed a series of smooth-particle hydrodynamics (SPH) experiments of circumbinary accretion disks with $(h/r) \in [0.02, 0.13]$.  They found an approximately linear suppression of the BHs' accretion rate with decreasing disk thickness for $(h/r) < 0.1$ and attributed this to decreasingly significant viscous torques.  However, \cite{Ragusa+2016} performed their parameter study of disk thickness at constant turbulent viscosity parameter $\alpha$ such that decreasing $(h/r)$ not only varies the magnitude of viscous torques, but also alters the magnitude of pressure gradients throughout the disk. They were also unable to accurately resolve the binary minidisks in their thinnest cases and did not calculate the binary torque or migration rate. \cite{Dorazio2016} also considered disks with Mach numbers between $3\leq \mathcal{M}\leq 30$, but focused on the impact of the temperature on the morphology of the disks (particularly on the transition from an annular ring to a round central cavity that occurs around a binary mass ratio of $q\equiv M_2/M_1 \approx 0.04$). They did not address the torques on the binary, or its orbital evolution.
 
The goal of this paper is to assess the effects of decreasing disk scale height (or equivalently, increasing Mach number) on binary accretion, binary-disk torques, and especially on the binary's orbital evolution. To our knowledge, this is the first numerical investigation of this dependence, which is of special interest, given that real AGN disks are thinner than the disks simulated so far in the literature.

This paper is organized as follows. In \S~\ref{sec:setup} we describe the details of our computational methodology and setup, as well as demonstrate numerical convergence.  In \S~\ref{sec:results} we present the results of our simulations and an analysis of the gravitational torques. Finally in \S~\ref{sec:conclusions} we summarize our main results and discuss some of their implications.

% ======================================================================
\section{Numerical methods}
\label{sec:setup}

% ======================================================================
\subsection{Simulation setup}

Our simulations were performed using the publicly available code \texttt{Mara3}, first described in \citet{Mara}. 
\texttt{Mara3} solves the vertically-averaged Navier-Stokes equations
\begin{align}
    \label{eq:NS1}
    &\frac{\partial \Sigma}{\partial t} + \mathbf{\nabla}\cdot(\Sigma\mathbf{v}) = \dot{\Sigma}_{\rm sink} \, , \\
    &\frac{\partial \Sigma \mathbf{v}}{\partial t} + \mathbf{\nabla } \cdot (\Sigma \mathbf{v}\mathbf{v} + P\,\mathbf{I} - \mathbf{T}_{\text{vis}}) = \dot{\Sigma}_{\rm sink} \mathbf{v} + \mathbf{F}_g
    \label{eq:NS2}
\end{align}
using a finite volume Godunov scheme in Cartesian coordinates with static mesh refinement. $\Sigma$ is the vertically integrated surface density of the disk, $\mathbf{v}$ is the gas velocity, and $P = \Sigma c_s^2$ is the vertically integrated gas pressure.  $\dot{\Sigma}_{\rm sink}$ is a mass-sink term which models the accretion of mass onto each black hole, and $\mathbf{F}_g = -\Sigma\,\nabla\phi$ is the vertically integrated gravitational force density, associated with the potential
\begin{align}
    \phi =\phi^{(1)} + \phi^{(2)} = -\frac{GM_1}{(r_1^2 + r_s^2)^{1/2}} - \frac{GM_2}{(r_2^2 + r_s^2)^{1/2}} \ .
\end{align}
Here $r_1$ and $r_2$ are the distances to each black hole and $r_s$ is the gravitational softening length which accounts for the vertical averaging of the gravitational force, and ensures the potential remains finite at the component positions. The sound speed $c_s$ is calculated according to a locally isothermal equation of state
\begin{equation}
    c_s^2 = -\phi / \mathcal{M}^2 \, ,
\end{equation}
where the Mach number is defined based on the vertically averaged thin-disk approximation,
\begin{align}
    \mathcal{M} \equiv v_\phi / c_s = (h/r)^{-1} \, .
\end{align}
The sound speed approaches $c_s^2 \propto M_i / r_i$ nearby either of the binary components, and $M / r$ far away from the binary. Here and throughout the text, $M = M_1 + M_2$ is the binary's total mass, and $r$ is the distance from the origin, coinciding with the binary's center of mass.

The mass accretion term in Eq. \ref{eq:NS1} and \ref{eq:NS2} is defined by a Gaussian kernel of size $r_{\rm sink}$ that removes each of the conserved quantities, $\mathbf{U} = (\Sigma, \Sigma\,v_x, \Sigma\,v_y)$, at a maximum rate $\tau_{\rm sink}$,
\begin{align}
    \dot{\mathbf{U}}_{\rm sink} = -\frac{\mathbf{U}}{\tau_{\rm sink}} \bigg( e^{-r_1^2\, /\, 2r_{\rm sink}^2} + e^{-r_2^2\, /\, 2r_{\rm sink}^2} \bigg) \ .
    \label{eq:sink}
\end{align}
The sink rate is chosen to be $\tau_{\rm sink}^{-1} = 8\,\Omega_{\rm b}$,
where $\Omega_{\rm b}$ is the binary's orbital angular frequency,
and the sink radius is set to be equal to the gravitational softening radius, $r_{\rm sink} = r_s = 0.05 a$, where $a$ is the binary separation.  The choice of $\tau_{\rm sink}$ does not significantly alter the results for circular orbits \citep{Moody19}.

The viscous term in Eq. (\ref{eq:NS2}) is given by the viscous stress tensor $\mathbf{T}_{\text{vis}}$.  The viscosity is chosen to be isotropic such that components of the tensor are 
\begin{align}
    T^{ij}_{\text{vis}} = \nu \,\Sigma \left( \frac{\partial v^i}{\partial x^j} + \frac{\partial v^j}{\partial x^i} - \frac{\partial v^k}{\partial x^k}\delta^{ij} \right) \, .
\end{align}
For this study we have selected a ``constant-$\nu$'' viscosity prescription, in which the kinematic viscosity coefficient is set to $\nu = \sqrt{2} \times 10^{-3} a^2 \Omega_{\rm b}$ globally and for all runs.\footnote{Except for a limited exploration of fixed $\alpha=0.1$ in \S~\ref{sec:results}.} This choice is made in favor of the more widely adopted $\alpha$ viscosity model \citep{Yike17, MML19, Moody19, Munoz19} in order to make high Mach number simulations computationally feasible. Indeed, in the $\alpha$-viscosity prescription the viscous time scale grows quadratically with Mach number, making it computationally prohibitive with available resources to run well-resolved simulations of globally relaxed high Mach number $\alpha$-disks. In contrast, with the constant-$\nu$ prescription, runs at different Mach numbers possess the same viscous time scale $t_\nu \sim r^2 / \nu$. $\nu$ and $\alpha$ are related by $\nu = \alpha c_s h$ \citep{SS1973}, so our $\mathcal{M}=10$ disk corresponds to the fiducial model with $\alpha = 0.1$ at $r = 2a$.
The effective-$\alpha$ as we raise the Mach number, thus, ranges from $\alpha \in [0.1, 1.6]$.

The initial condition is a quasi-steady disk\footnote{"Quasi-steady" here and throughout this paper refers to the fact that the disk is never in a true steady-state because of its viscous expansion, but nevertheless, it relatively quickly settles to a constant angular momentum transfer rate (see Figure \ref{fig:dots}).} of finite extent with peak density at $r_d = 4\,a$ and a mildly depleted cavity region,
\begin{eqnarray}
    \Sigma &=& \Sigma_0 \, e^{-(r/r_{d} - 1)^2 / 2} + \Sigma_{\rm out} \\
    \mathbf{v} &=& \left( \frac{GM}{r} + \frac{r}{\Sigma} \frac{\partial P}{\partial r} \right)^{1/2}\,\hat{\phi} \, .
\end{eqnarray}
Here $\Sigma_{\rm out} = 10^{-10} \Sigma_0$ is a constant ambient density floor. The initial condition is a steady-state solution to Equations \ref{eq:NS1} and \ref{eq:NS2} with zero viscosity ($\nu = 0$) and a central gravitational potential. Viscous drift/diffusion develops self-consistently as the disk relaxes.

All simulations are performed with an equal-mass binary on a fixed circular orbit. Because the mass of the inner region of AGN disks is typically far smaller than their central black holes, it is well motivated to ignore the disk's self-gravity, as well as the evolution of the binary orbit during the simulated 1000 binary orbits.  Moreover, it has been found that orbits with a small amount of eccentricity ($\epsilon \lesssim 0.1$) have their eccentricity damped and approach the circular limit \citep{MML19, Zrake2020}.
Figure \ref{fig:reliefs} depicts sample simulations, showing the disk surface density at four Mach numbers, after approximately a viscous time, out to $r = 8a$ (the full domain extends out to 32a). Each run has developed the standard characteristics of near-equal-mass binaries accreting from a CBD --- namely minidisks surrounding each black hole, and an eccentric cavity characterized by an $m=1$ surface density feature near the cavity wall, referred to as a ``lump'' \citep{2008ApJ...672...83M, Shi+2012,Roedig2012,Dorazio2013,Dorazio2016,   Farris15, Yike17}. Note that as $\mathcal{M}$ is increased, the cavity develops more complex structures, and the lump becomes sharper and denser, something that was also pointed out in \cite{Ragusa+2016}. We discuss the implications of this trend in \S~\ref{sec:results} below.

% ======================================================================
\subsection{Simulation diagnostics}

\begin{figure*}
        \includegraphics[width=\columnwidth]{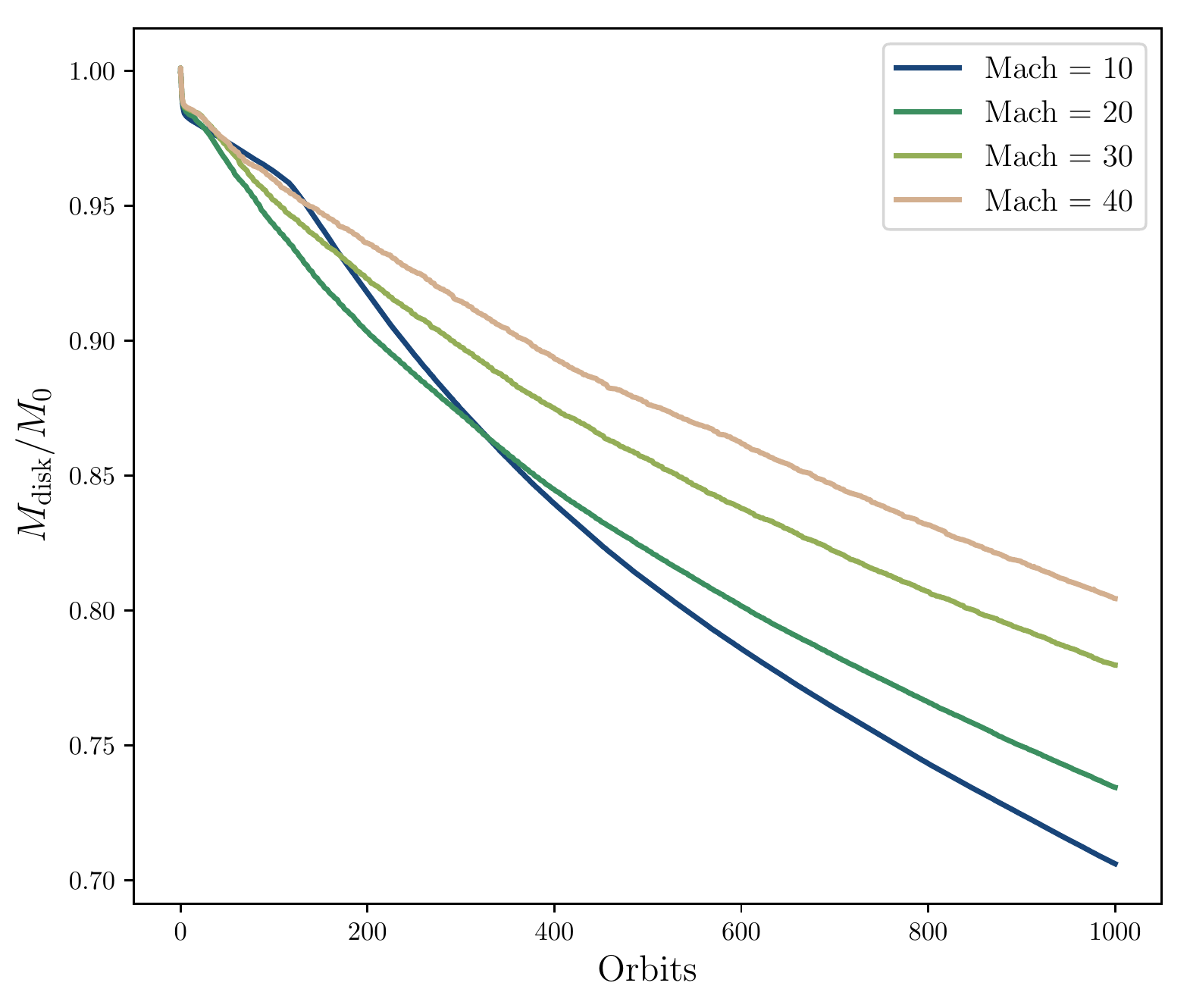}
        \includegraphics[width=\columnwidth]{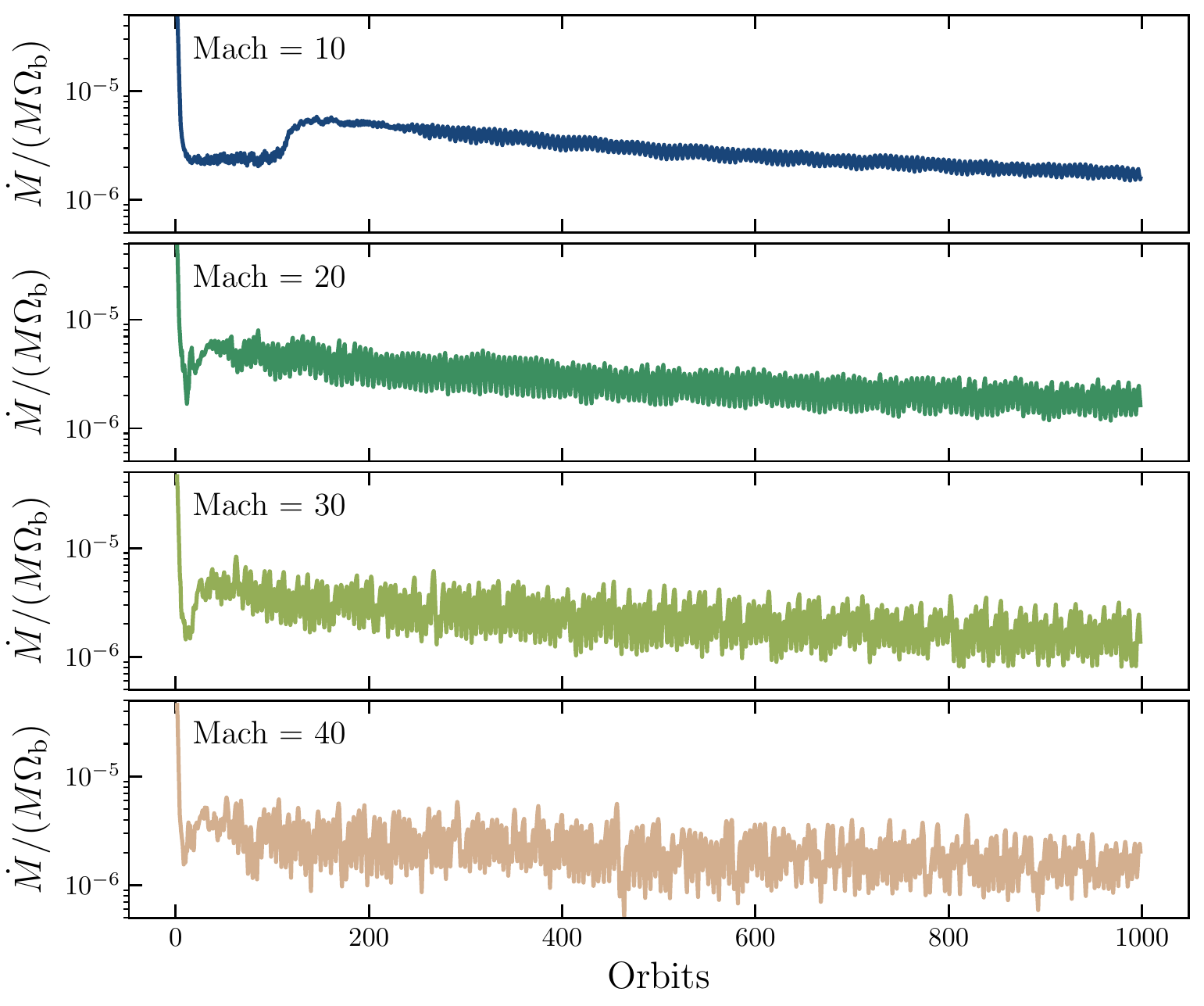}
        \protect\caption{Left panel: The mass of the disk is gradually depleted as the binary accretes, but higher $\mathcal{M}$ disks accrete more slowly. Right panel: $\dot{M}$ reaches a quasi-steady-state with a long-term secular decrease from the depletion through the sinks and viscous spreading 
        of the disk for all $\mathcal{M}$.}
     \label{fig:disk_mass}
\end{figure*}

The goal of our simulations is to predict the evolution of the binary separation $\dot a$ induced by gas accretion onto the black holes. 
Assuming that the binary remains on a circular orbit, and maintains its mass ratio, this ``migration rate'' depends only on the rate of change of angular momentum and mass (with the change in binding energy fixed by the circularity requirement), 
and is given analytically by
\begin{equation}\label{eq:adot}
    \frac{\dot a}{a} = 2 \left(\frac{\ell_0}{\ell} - \frac{3}{2} \right) \frac{\dot M}{M} \, ,
\end{equation}
where $\ell \equiv L / M$ is the specific angular momentum of the binary, $\ell_0 \equiv \dot L / \dot M$ is the so-called accretion eigenvalue, $\dot L$ is the total torque exerted on the binary, and $\dot M$ is the total mass accretion rate onto the binary. 
\footnote{Equation \ref{eq:adot} is equivalent to Equation 39 in \cite{MML17}. However we have chosen to define $\ell$ as the binary specific angular momentum (i.e. angular momentum per \emph{total binary mass}), which matches the specific angular momentum of gas co-orbiting with the BHs for an equal-mass binary.  This is 4 times smaller than the specific angular momentum in the equivalent one-body problem, $\sqrt{G M a}$, which is defined as angular momentum per \emph{reduced mass}.  The latter definition was used by \cite{MML17}, and also appears in \cite{MML19} and \cite{Moody19}.}
Equation \ref{eq:adot} is readily obtained by differentiating in time the expression for the binary orbital angular momentum $L = \frac{1}{4}\sqrt{G M^3 a}$. It also makes clear that the binary expands when the accretion eigenvalue $\ell_0$ is greater than a critical value $\ell_c \equiv \frac{3}{2} \ell$ ($= \frac{3}{8} \sqrt{GMa}$) and the binary shrinks when $\ell_0 < \ell_c$. In other words, outward migration requires that each parcel of accreted gas delivers on average at least 50\% more specific angular momentum than that of the binary. $\ell_0 > \ell_c$ implies that the torque applied by the accreting gas overcomes the orbital hardening associated with the increasing binary mass.

Measurements of the migration rate reported in \S~\ref{sec:results} are obtained by inserting the simulation-computed time series $\dot M$ and $\dot L$ into Equation \ref{eq:adot}. \texttt{Mara3} is configured to compute these time series in a conservative fashion, such that the total mass removed by the sink term $\dot \Sigma_{\rm sink} dA$ during each time step $\Delta t$ exactly equals $\dot M \Delta t$,
\begin{align}
    \Delta M = \Delta t \int \dot{\Sigma}_{\rm sink} \, dA \ .
    \label{eq:mdot}
\end{align}
Similarly, the angular momentum impulse delivered to the black holes in a time step $\Delta t$ is precisely the angular momentum $\Delta L$ removed from the gas in that time interval. The increments $\Delta M$ and $\Delta L$ are then time-integrated according to the same Runge-Kutta stepping as is used to advance Equations \ref{eq:NS1} and \ref{eq:NS2}.

The total torque $\dot{L} = \dot{L}_{\rm grav} + \dot{L}_{\rm acc}$ consists of the gravitational torque $\dot{L}_{\rm grav}$ on the binary, and the rate $\dot{L}_{\rm acc}$ of angular momentum consumed directly through the sinks. The change $\Delta L = \dot L \Delta t$ of binary angular momentum in each time step $\Delta t$ is computed according to
\begin{align}
    \Delta L_{\rm grav} &= \hat{z} \cdot \left( \mathbf{r}^{(1)} \times   \Delta \mathbf{p}_{\rm grav}^{(1)} + \mathbf{r}^{(2)} \times \Delta \mathbf{p}_{\rm grav}^{(2)} \right) \nonumber \\
    \Delta L_{\rm acc} &= \hat{z} \cdot \left( \mathbf{r}^{(1)} \times \Delta \mathbf{p}^{(1)}_{\rm acc} + \mathbf{r}^{(2)} \times \Delta \mathbf{p}^{(2)}_{\rm acc} \right) \, , \nonumber
    %\label{eq:ldots}
\end{align}
where the four linear impulse terms are computed in a conservative fashion as mentioned previously,
\begin{align}
    \label{eq:pacc}
       \Delta \mathbf{p}^{(i)}_{\rm acc } &= - \Delta t \int \dot \Sigma_{\rm sink}^{(i)} \mathbf{v} \, \mathrm{d}A \nonumber \\
       \Delta \mathbf{p}^{(i)}_{\rm grav} &= - \Delta t \int \mathbf{F}^{(i)}_g \, dA \, \, ; \qquad \mathbf{F}^{(i)}_g = -\Sigma \nabla \phi^{(i)} \, . \nonumber
    %\label{eq:pgrav}
\end{align}
%

% ======================================================================
\subsection{Resolution and convergence}

\texttt{Mara3} employs block-structured static mesh refinement in a nested-box topology in order to concentrate numerical resolution on the minidisks and inner cavity. The computational domain extends from $-R$ to $R$ in both directions. The mesh blocks are square, with $n_c$ zones per side, and are refined by factors of 2 up to a maximum depth $d$ such that the finest grids have mesh spacing
\begin{align}
    \Delta r_{\rm min} = \frac{R}{2^{d-1}\, n_c} \, .
\end{align}
Simulations reported here all have $R = 32a$ and $d = 6$. In order to establish numerical convergence, we have performed simulations with $n_c = \{32, 64, 96\}$, corresponding to resolutions inside the binary cavity of $\Delta r_\text{min} = \{0.0312 a, 0.0156 a, 0.0104 a\}$ respectively. Figure \ref{fig:disk_mass} shows the evolution of disk mass and accretion rate for Mach numbers $\mathcal{M} = \{10, 20, 30, 40\}$. The time-averaged accretion rate $\langle \dot{M} \rangle$ and torque eigenvalue $\langle\dot{L}\rangle / \langle\dot{M}\rangle$ for each resolution case are shown in Figure \ref{fig:mach_study}. The higher-resolution runs with $\Delta r = 0.0156 a$ and $\Delta r = 0.0104 a$ are consistent with one another to within $10\%$ for $\mathcal{M} = 10 - 30$, with slightly larger deviations for $\mathcal{M} > 30$. Simulations presented in \S~\ref{sec:results} were performed with $n_c = 64$ ($\Delta r_\text{min} = 0.0156 a$), and run through 1000 binary orbits.  Although these runs deviate more significantly from the highest-resolution suite at the highest Mach number ($\mathcal{M} = 40$), this deviation does not influence the conclusions of this work with regard to the transition occurring at $\mathcal{M}\sim 25$.

% ======================================================================% ======================================================================
\section{Results and discussion}
\label{sec:results}

The main results of our study are summarized in Figure \ref{fig:mach_study}. The top panel shows the accretion rate onto the binary, averaged between 300 and 500 orbits, and the bottom panel shows the averaged angular momentum transfer per unit accreted mass (the accretion eigenvalue $\ell_0$). We observe modest suppression of the accretion rate with higher $\mathcal{M}$, even though these models were run with the same kinematic viscosity $\nu = \sqrt{2} \times 10^{-3} a^2 \Omega_{\rm b}$ (note that a similar effect was seen by \cite{Ragusa+2016}). At $\mathcal{M}=10$, the accretion eigenvalue is greater than the critical value $\ell_c$, corresponding to binary expansion, and in qualitative agreement with other studies based on $\mathcal{M}=10$ disks \citep{MML19, Moody19, Munoz19}. However we observe a systematic reduction of $\ell_0$ as the Mach number is increased, with $\ell_0$ becoming smaller than $\ell_c$, corresponding to binary inspiral, at $\mathcal{M} \sim 25$, and smaller still for larger $\mathcal{M}$. If this trend continues to Mach numbers $\gtrsim 100$, then the very thin disks in AGN are predicted to rapidly drive a SMBHB toward coalescence.

Additionally, we performed a limited exploration of varying the Mach number at constant $\alpha=0.1$, equivalent to our fiducial constant-$\nu$ run at $\mathcal{M}=10$ for $r=2a$.  These were run out to viscous times corresponding to 300 orbits in the fiducial run (into the regime of constant $\ell_0$; see Figure \ref{fig:dots}), and the result of this exercise is shown by the red dashed curve in the bottom panel of Figure ~\ref{fig:mach_study}. It suggests that $\ell_0$ --- and therefore, the primary finding of this paper --- is not significantly altered in this case. In future work we intend to explore this issue more thoroughly.

The binary migration rate, as computed from Equation \ref{eq:adot} and the data in Figure \ref{fig:mach_study}, is shown in Figure \ref{fig:delta_a} for the two highest resolution runs (over the same $300-500$ orbit window). The migration rate becomes negative at and above $\mathcal{M} \sim 25$, consistent with where the accretion eigenvalue becomes smaller than $\ell_c$. Figure \ref{fig:delta_a} also shows the relative contributions of gravitational forces and accretion to the total migration rate. We observe that the contribution to $\dot a$ from the accretion of mass and angular momentum is relatively insensitive to the Mach number; the downward trend in $\dot a$ is due to a systematic reduction of the gravitational torque with increasing Mach number. The physical mechanism for this is explored  in \S~\ref{sec:torque} below.

% ======================================================================
\subsection{Finite disks}
\label{sec:finite_disks}

\begin{figure}
    \includegraphics[width=\columnwidth]{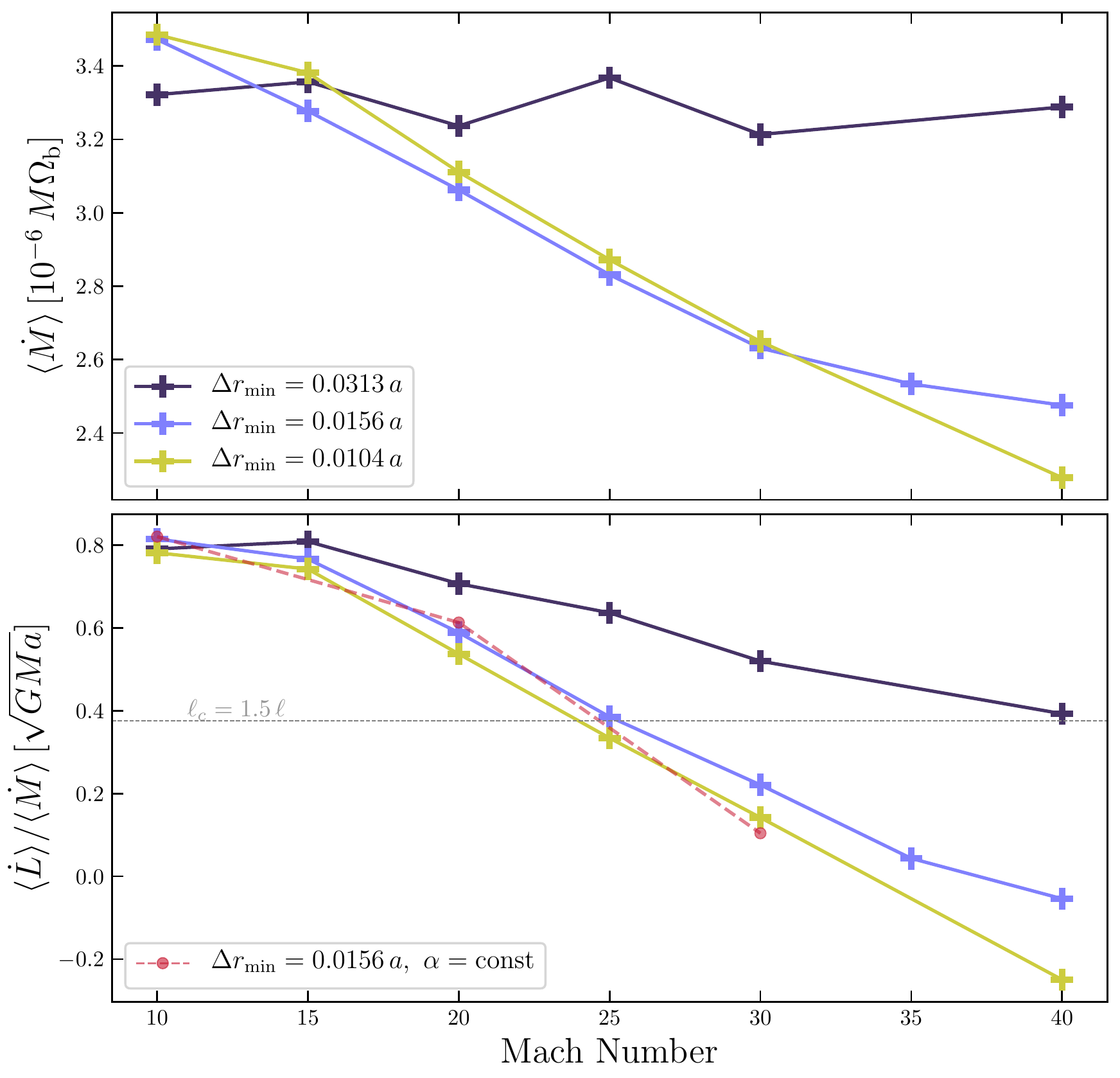}
    \protect\caption{
    Average accretion rate and average change in angular momentum per unit accreted mass as functions of Mach number at three different resolutions. Each quantity decreases with increasing Mach number. 
    Most notably, the angular momentum gained falls below the critical threshold $\ell_c$ (shown as the gray horizontal line) and binary evolution therefore switches signs between $\mathcal{M} = 20 - 30$, implying that at high Mach numbers, the binary-disk interaction drives binaries towards merger.  The results show good convergence for $\Delta r_{\rm min} = 0.0156\, a$, except at the highest Mach number.  The red-dashed-line denotes a limited study at constant $\alpha$ and suggests that our primary conclusion is not strongly dependent on the choice of viscosity.}
    \label{fig:mach_study}
\end{figure}

\begin{figure}
    \includegraphics[width=0.45\textwidth]{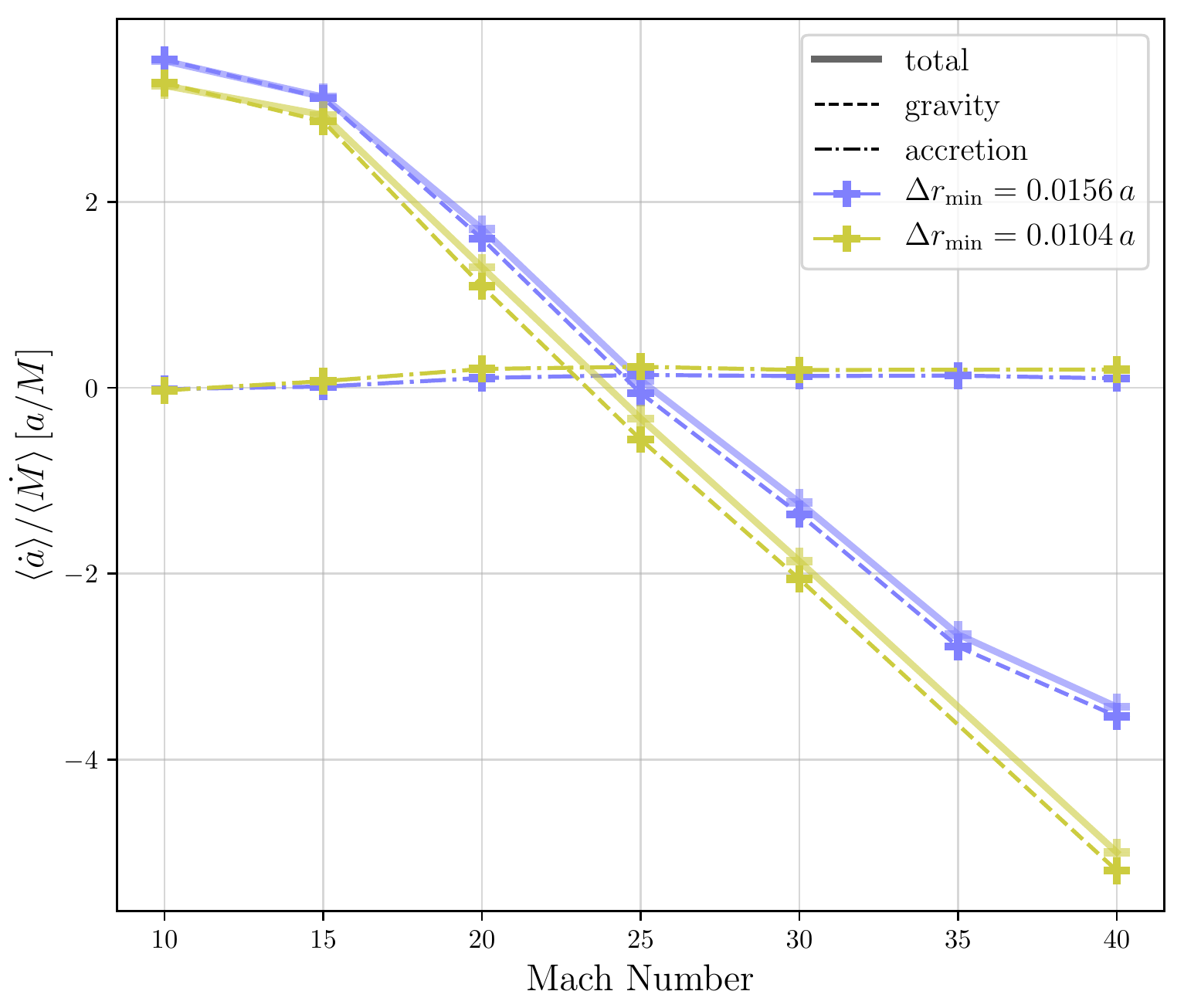}
    \protect\caption{Change in binary separation computed from Eq.~\ref{eq:adot}, decomposed into its gravitational and accretion components for the two higher resolutions in the considered window (300-500 orbits).  Besides the transition from outspiraling to inspiraling binaries at high $\mathcal{M}$, we also find that the change in binary separation due to accretion is approximately constant for all Mach numbers considered. Therefore, the transition to in-spiraling binaries at high $\mathcal{M}$ is due to decreasing gravitational torques.}
    \label{fig:delta_a}
\end{figure}

\begin{figure}
    \includegraphics[width=0.99\columnwidth]{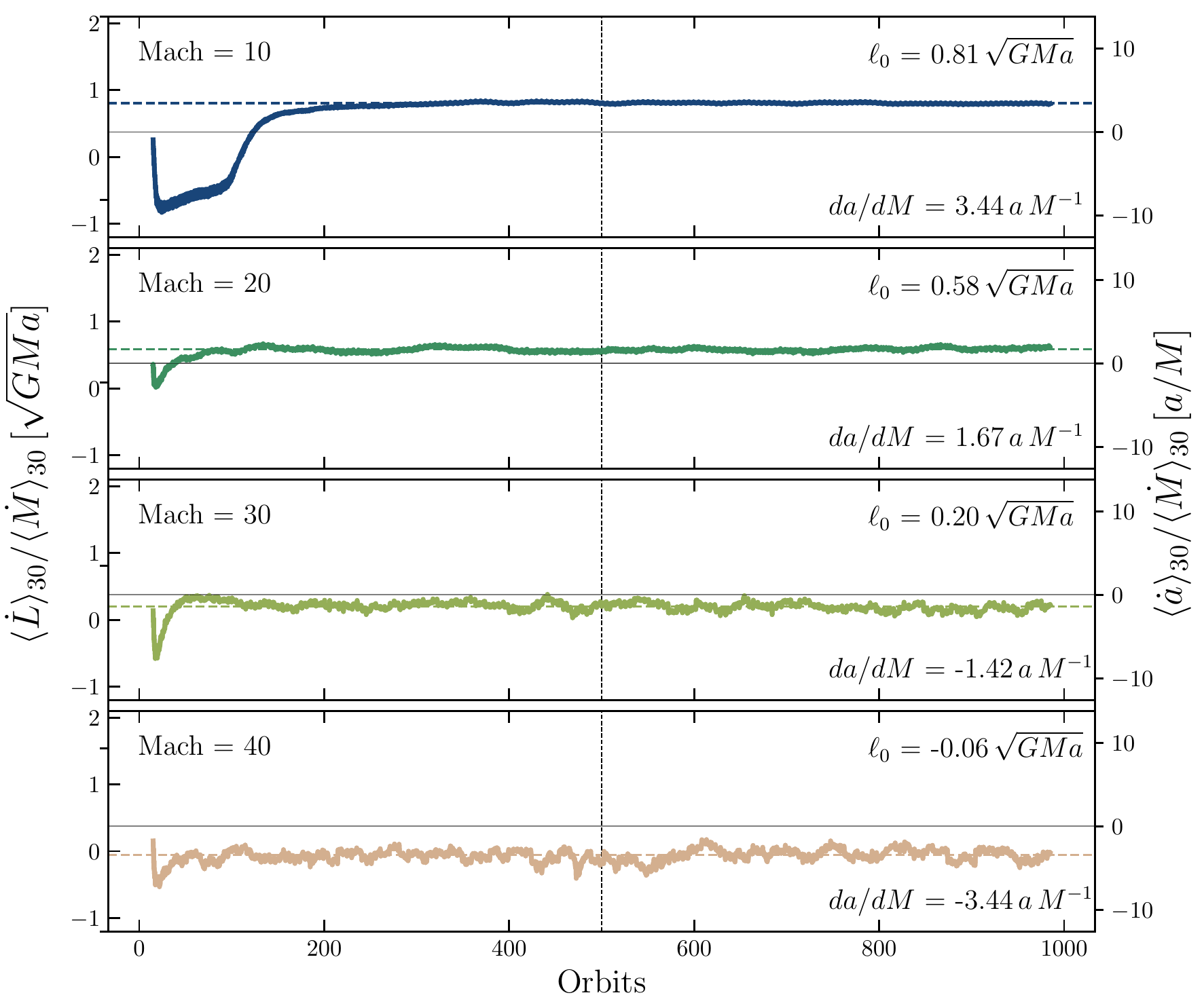}
    \protect\caption{30-orbit averages for $\dot{L}$ and $\dot{a}$ normalized by the 30-orbit average of $\dot M$ for $\mathcal{M} = \{10, 20, 30 , 40\}$. The average between 500 and 1000 orbits, $\ell_0$ and $da/dM$, are shown by the horizontal dashed lines. $\langle \dot{L} \rangle_{30} / \langle \dot{M} \rangle_{30}$ remains constant after a  transient initial phase despite the secular evolution in disk structure and accretion rate.  Most notably, the angular momentum gained per unit accreted mass crosses the threshold $\ell_c=\frac{3}{8} \sqrt{GMa}$ (gray horizontal lines) to result in inspiraling binaries ($da/dM < 0.0$) between $\mathcal{M} = 20-30$.  Variability also grows with Mach number, but no apparent long-term trends emerge.}
    \label{fig:dots}
\end{figure}

We have considered disks of finite mass and extent. These disks viscously expand into the low-density ambient medium, such that a precise steady-state is never attained. Nevertheless, we find that the transfer of angular momentum always occurs at a constant fraction of the accreted mass. In Figures \ref{fig:disk_mass} and \ref{fig:dots} we show that as the circumbinary disk is depleted (and $\dot M$ slowly decreases), both the accretion eigenvalue $\langle\dot{L}\rangle / \langle\dot{M}\rangle$ and consequently the orbit-averaged migration rate $\langle\dot{a}\rangle / \langle\dot{M}\rangle$ remain constant throughout the quasi-steady simulation phase ($\gtrsim 200$ orbits). The left panel of Figure \ref{fig:disk_mass} shows the total disk mass versus time at four Mach numbers, and the right panel shows the accretion rate through 1000 orbits. Figure \ref{fig:disk_mass} also confirms the slow secular evolution of $\dot M$ as the disk is depleted, and is consistent with Figure \ref{fig:mach_study} in that $\dot M$ is modestly suppressed as $\mathcal{M}$ is increased. Through 1000 orbits, the disk has lost about 30\% of its initial mass in the $\mathcal{M} = 10$ case, while at $\mathcal{M} = 30$ the disk has only been depleted by about 20\%. Additionally, we have confirmed that this mass is lost through the sinks and not through the outer boundary.

Figure \ref{fig:dots} shows the time series for the accretion eigenvalue calculated over 30-orbit windows, $\langle \dot{L} \rangle_{30} / \langle \dot{M} \rangle_{30}$. The associated change in binary separation per unit accreted mass $\langle \dot{a} \rangle_{30} / \langle \dot{M} \rangle_{30}$ is shown on the right vertical axis. The accretion eigenvalue $\ell_0$ and the migration rate per accreted mass $da / dM$ corresponding to the whole quasi-steady evolution phase ($t > 500$) are shown as horizontal lines. The angular momentum gained per unit accreted mass crosses the threshold $\ell_c = \frac{3}{8} \sqrt{GMa}$  to result in inspiraling binaries ($da/dM < 0.0$) between $\mathcal{M} = 20-30$. We find that the accretion eigenvalue is steady over hundreds of orbits, in agreement with \cite{Munoz19} for finite $\mathcal{M} = 10$ disks, and the accretion eigenvalue $\ell_0 = 0.81 \sqrt{G M a}$ is in reasonable quantitative agreement with the value $\approx 0.7 \sqrt{G M a}$ reported there. This quasi-steady behavior appears to apply equally well to the high Mach number runs. The amount of variability in the instantaneous $\dot L / \dot M$ is seen to increase with Mach number, but none of the runs show secular evolution of $\dot L / \dot M$ over time.

\begin{figure}
    \includegraphics[width=\columnwidth]{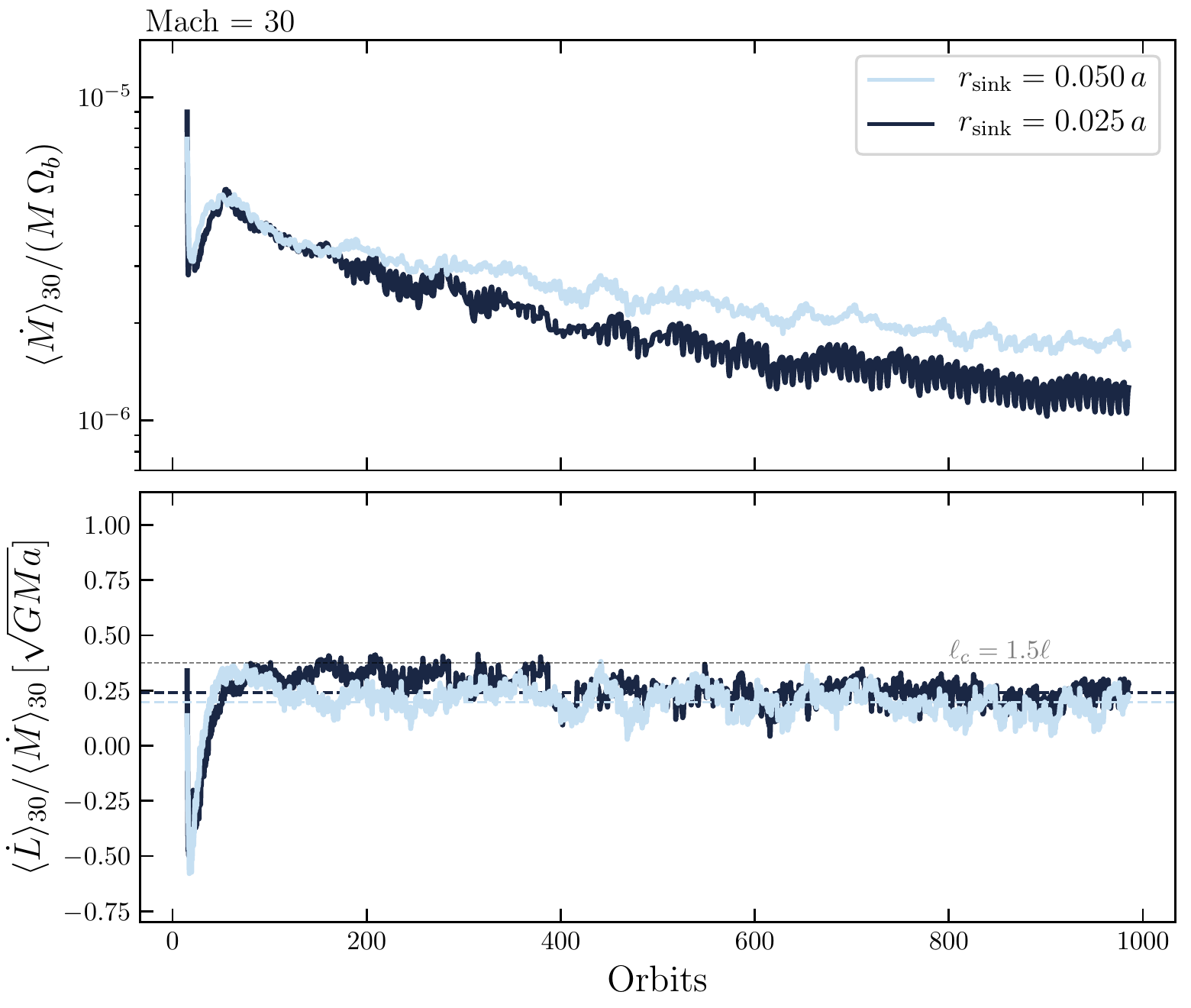}
    \protect\caption{30-orbit averages for accretion rate and eigenvalue for two different sink sizes.  The fiducial size is $r_{\rm sink} = 0.05$. Accretion rate decreases with sink size while accretion variability grows.  The torque per unit accreted mass, however, appears mostly indifferent to the halving of the sink radius.}
    \label{fig:sink}
\end{figure}

We have confirmed that our measurement of the migration rate is not dependent on our choice of sink radius by repeating the $\mathcal{M} = 30$ run with $r_{\rm sink} = 0.025 a$, i.e. reduced to half of its fiducial value. The results from this run are shown in Figure \ref{fig:sink}. The smaller sink run has marginally smaller $\dot M$, but the large and small sink runs have essentially identical $\ell_0$.

\begin{figure*}
    \includegraphics[width=\columnwidth]{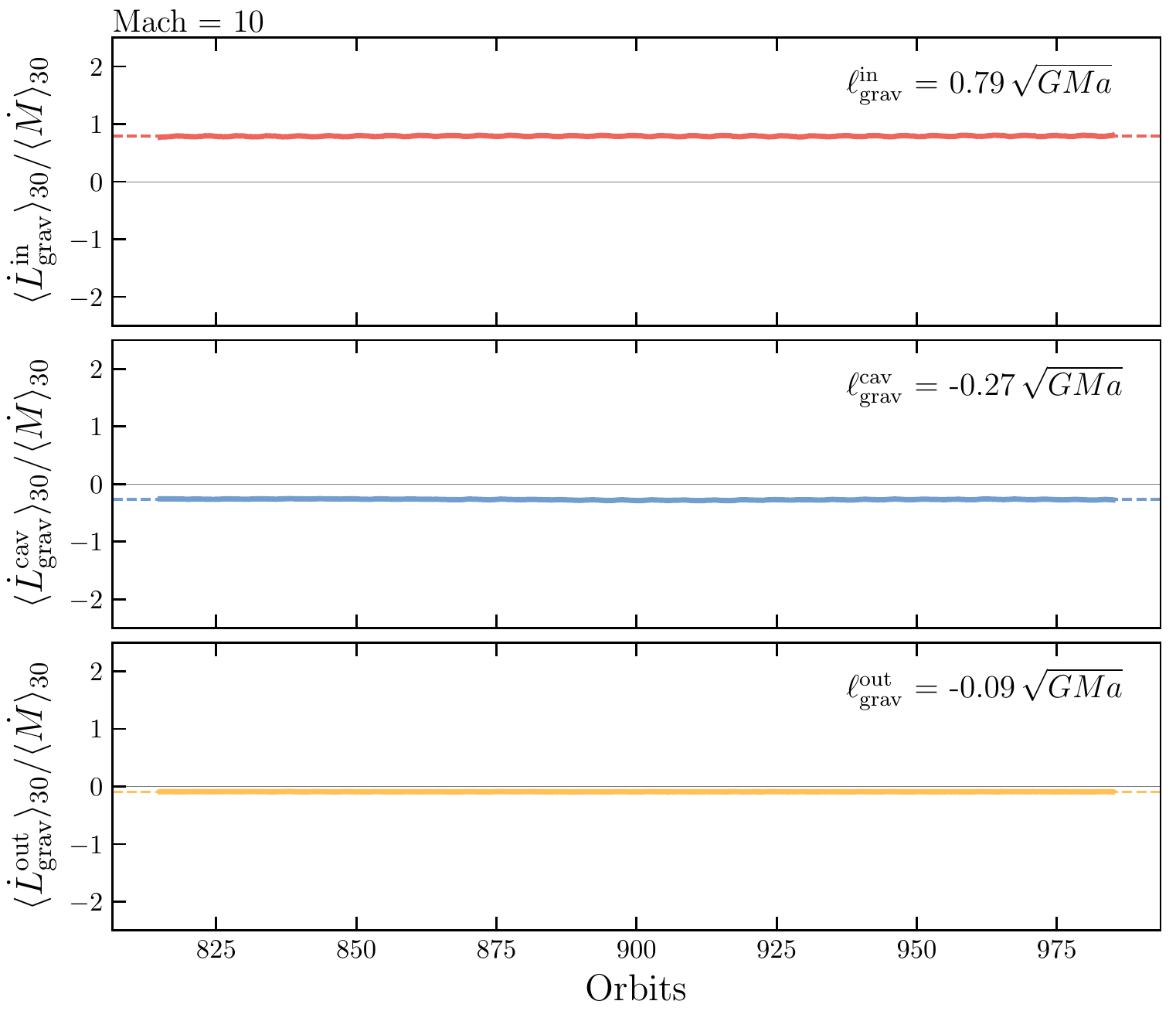}
    \includegraphics[width=\columnwidth]{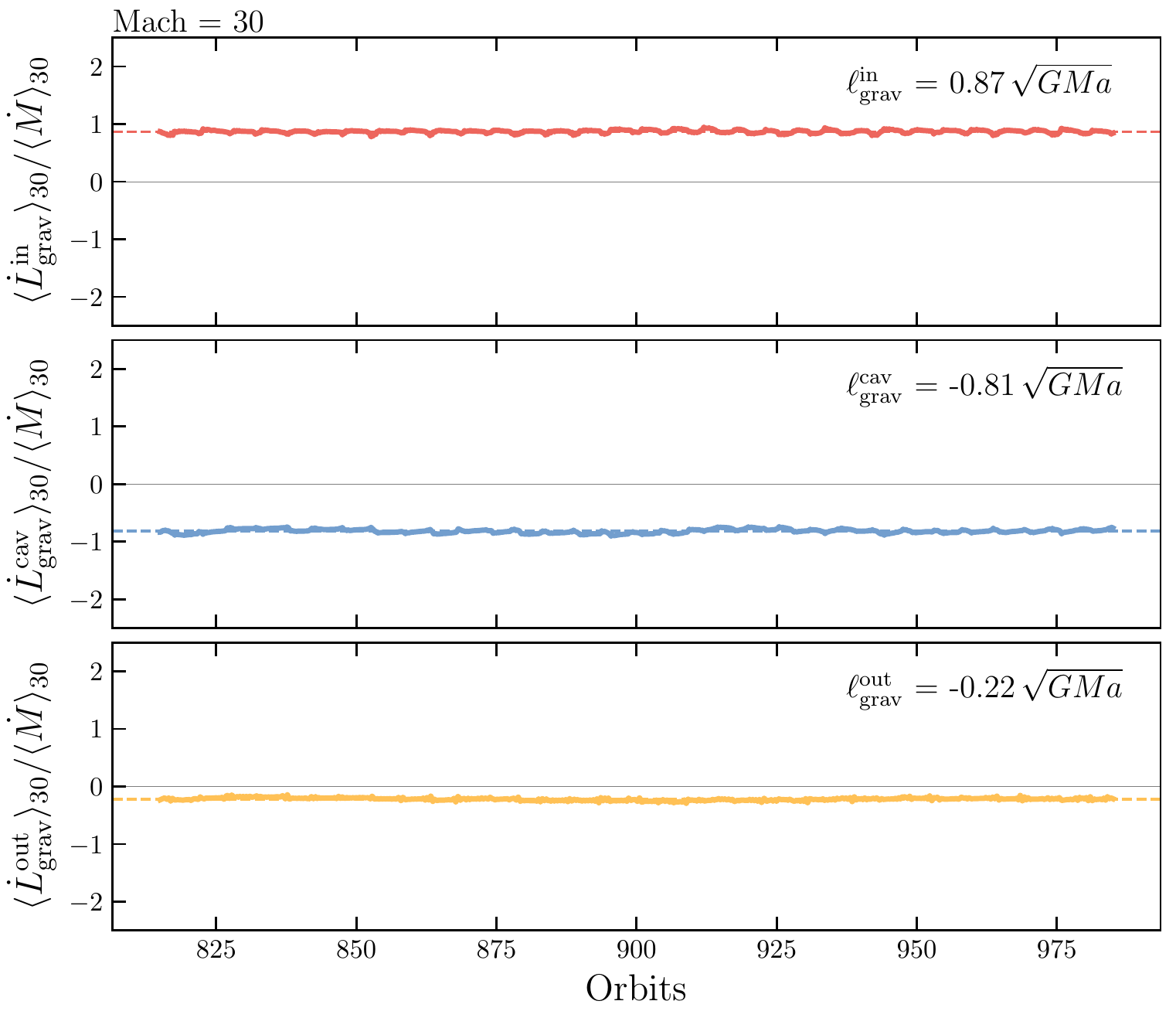}
    \protect\caption{30-orbit averages of the gravitational torques from three distinct regions in the circumbinary disk: $r < a$ (in, top panels)
    , $a < r < 2.5a$ (cav, middle panels), $r > 2.5a$ (out, bottom panels). The full average, $\ell^j_{\rm grav}$, is shown by the horizontal dashed line and its value reported in the top right corner. Torques in the outer disk (out) are negative in accordance with theory, but for $\mathcal{M} = 10$ the positive torques in the inner-most region (in) dominate the outer torques.  At $\mathcal{M} = 30$
    , however, the outer torque, and most notably, the torque from the cavity region (cav) are significantly more negative, resulting in net-negative gravitational torques.}
    \label{fig:torque_series}
\end{figure*}

\begin{figure}
    \includegraphics[width=\columnwidth]{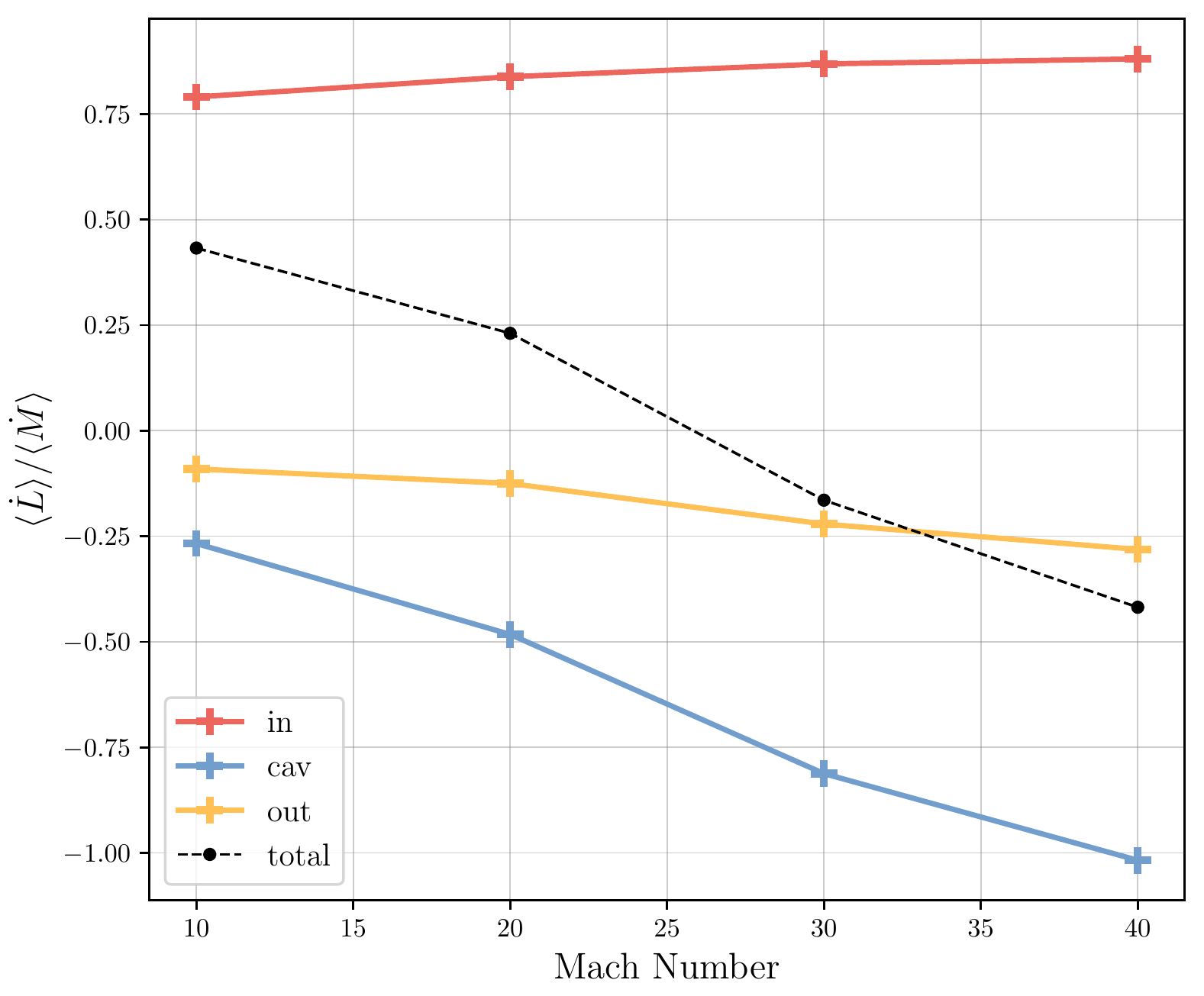}
    \protect\caption{Decomposed gravitational torques as a function of Mach number. The growth in torque magnitude for the inner (in) and outer (out) regions ($r < a$ and $r > 2.5a$ respectively) roughly offset. However, torques from near the cavity wall (cav, $a<r<2.5a$) become nearly 4 times more negative and make the total torque on the binary negative, driving it toward merger at high Mach number.}
    \label{fig:segmented}
\end{figure}

\begin{figure*}
    \includegraphics[width=\textwidth]{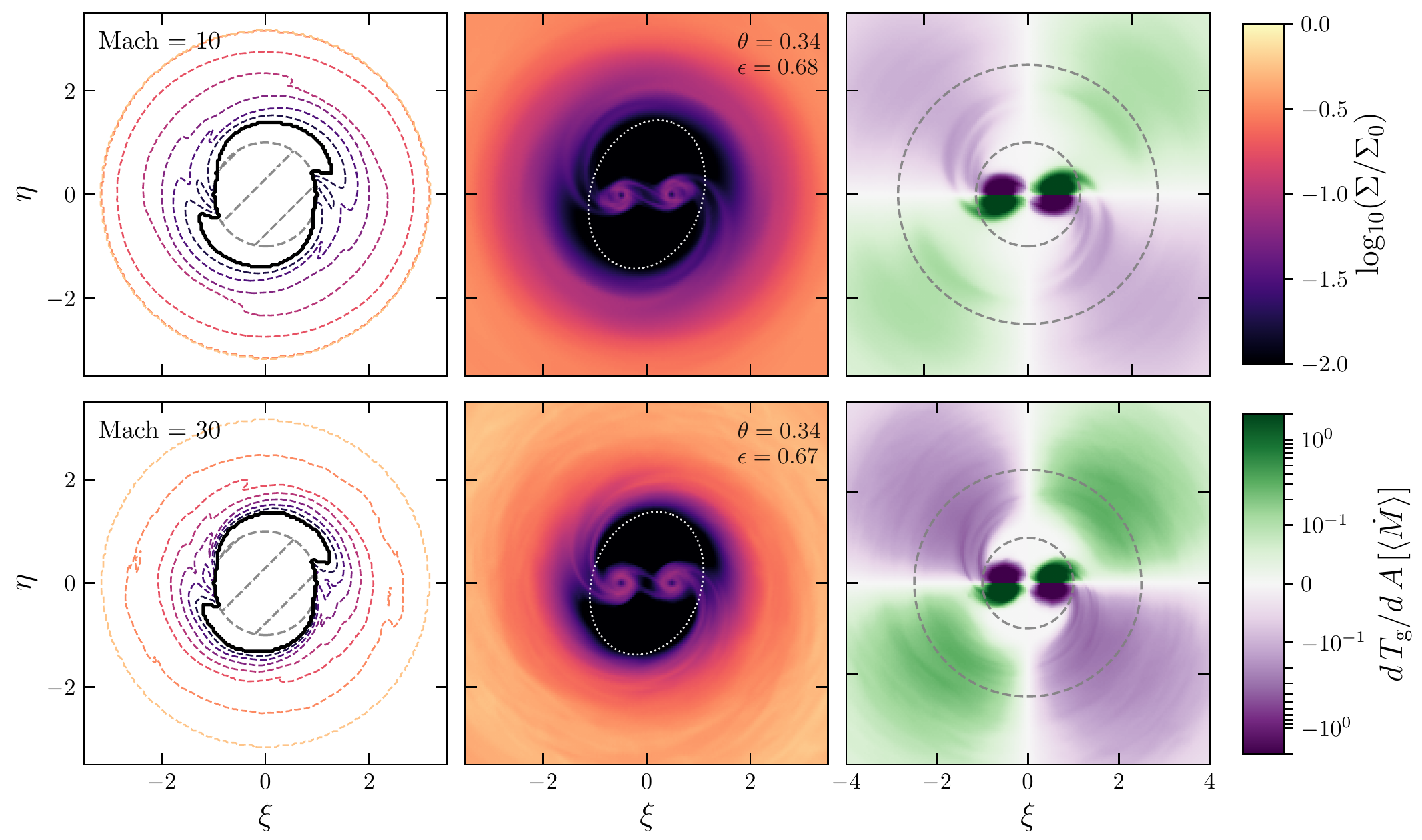}
    \protect\caption{Maps of the surface density (middle panels) and gravitational torque density (right panels) in the binary center of mass frame averaged over 50 orbits. The top and bottom rows correspond to $\mathcal{M} = 10$ and 30, respectively. The dashed circles at $r_{\rm in} = a$ and $r_{\rm cav} = 2.5 a$ in the torque density plot delineate the three zones considered in Figures \ref{fig:torque_series} and \ref{fig:segmented}. The $\mathcal{M} = 30$ case shows a comparative increase in surface density in the cavity region and the inner portions of the outer disk.  This is reflected in an increased torque magnitude in the cavity region, $a < r < 2.5 a$.  Iso-density contours are drawn from the surface density data in the left column and the solid black contour defines the time-averaged disk cavity.  The dotted white ellipse is the best-fit ellipse to this ``cavity'' contour and its rotation angle ($\theta$, in radians from the vertical) and eccentricity ($\epsilon$) are reported in the top right corners.  The time-averaged shape and orientation of the central cavity is similar for all $\mathcal{M}$ we considered.}
    \label{fig:density_fields}
\end{figure*}

% ======================================================================
\subsection{Torque distributions}
\label{sec:torque}

Figure \ref{fig:delta_a} shows that the change in binary separation due to accretion is insensitive to $\mathcal{M}$, while the gravitational torque becomes increasingly negative at higher Mach number, and is responsible for the transition to inward migration. Here we address the location in the disk where the increasingly negative gravitational torque is coming from.

To identify the source of the negative gravitational torque, we follow \cite{Yike17} and \cite{MML19} and compute the gravitational torque per accreted mass, $\ell_{\rm grav} = \dot{L}_{\rm grav} / \dot{M}$, in three different annuli: (1) the innermost region $r < a$, (2) the cavity region $a < r < r_{\rm cav}$, and (3) the outer region $r > r_{\rm cav}$, where we adopt $r_{\rm cav} = 2.5 a$.  These diagnostics are output by the code at the same cadence as the other time series quantities. Figure \ref{fig:torque_series} shows the 30-orbit average of these localized torques as a function of time, between 800 and 1000 binary orbits. For $\mathcal{M} = 10$ (shown in the left panel), the positive torque from the innermost region ($r < a_0$) is roughly twice the magnitude of the negative torques from the cavity and outer disk regions. This behavior is consistent with \cite{MML19}. However, for the $\mathcal{M} = 30$ case, (shown in the right panel), we find that the torque from the cavity region is significantly more negative than at Mach 10, while the torque magnitude from the inner and outer regions is only slightly larger.

To see this more clearly, we show in Figure \ref{fig:segmented} the 200-orbit window average of the gravitational torque per unit accreted mass, from each of the three regions, and at each Mach number $\mathcal{M} = \{10, 20, 30, 40\}$. While the positive torque in the innermost region ($r < a$) increases by only $\sim 10\%$, the negative torque from the cavity region ($a < r < 2.5 a$) is amplified by a factor of almost 4.

\begin{figure}
    \includegraphics[width=\columnwidth]{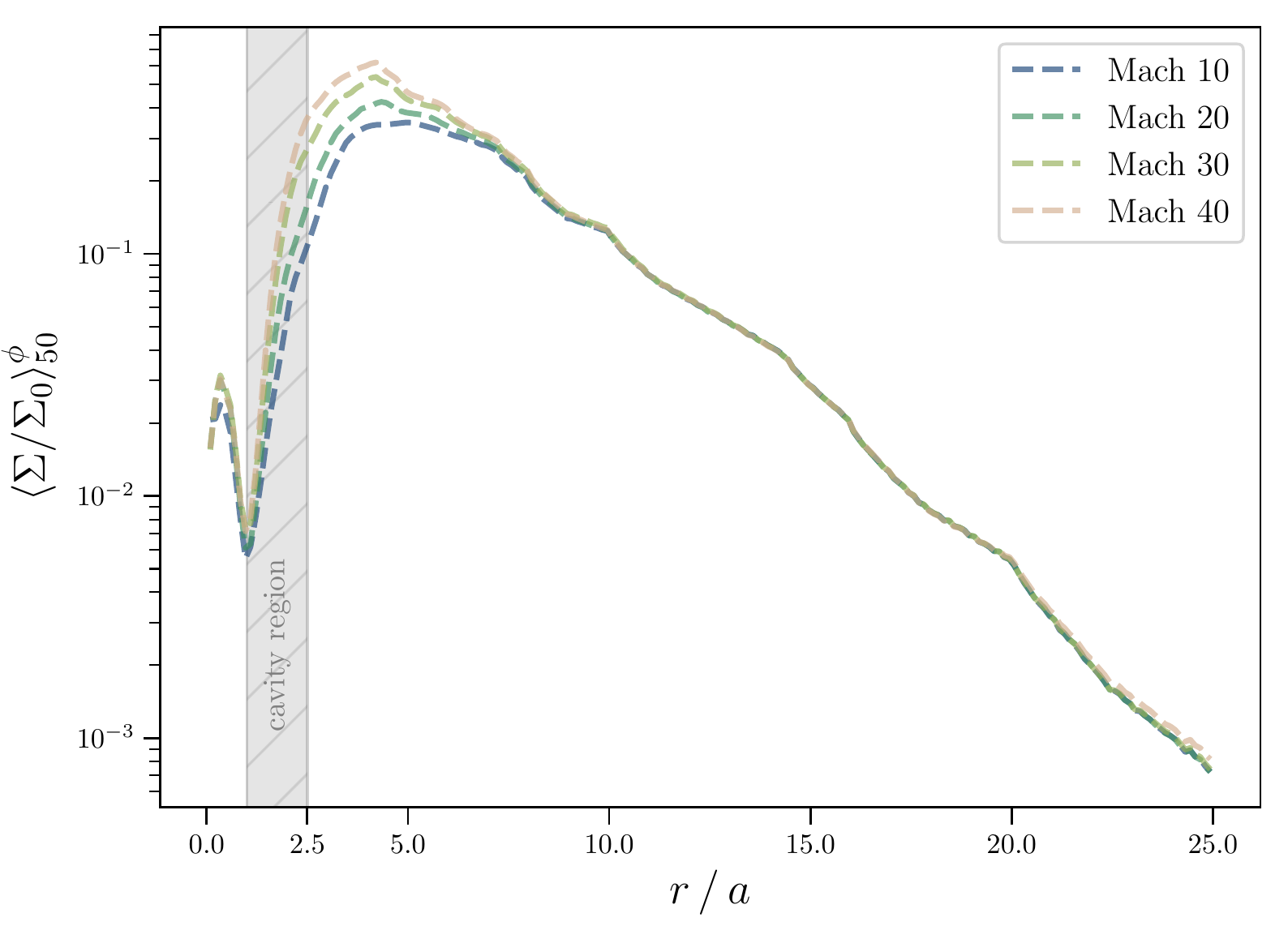}
    \protect\caption{Time and azimuthally averaged density profiles for $\mathcal{M} = \{10, 20, 30, 40\}$ ($\langle \cdot \rangle_{50}^\phi$ denotes a 50-orbit average and an average over $\phi$). We see that, consistent with Figure \ref{fig:density_fields}, the peak density near the cavity wall grows with $\mathcal{M}$. However, the density in the mini-disks at $r = 0.5a$ remains nearly constant with only a slight increase with $\mathcal{M}$ and the outer disk structure is identical for all $\mathcal{M}$. The cavity region ($a < r < 2.5a$) is shaded for reference.}
    \label{fig:sig_profile}
\end{figure}

To further examine the origin of the gravitational torque, we have reconstructed the two-dimensional torque density distribution from the surface density,
\begin{align}
    \frac{dT_{\rm g}}{dA} = \hat{z} \cdot \bigg( \mathbf{r}^{(1)} \times \mathbf{F}_g^{(1)} +  \mathbf{r}^{(2)} \times \mathbf{F}_g^{(2)} \bigg) \ .
\end{align}
We output snapshots of the surface density field 10 times per orbit and rotate each of these by the orbital phase $\theta_b$, so that the separation vector lies along the $x$-axis. The rotated coordinates have axes $\xi$ and $\eta$, and are obtained via the area-preserving transformation
$x = a\, (\xi cos{\theta_b} - \eta \sin{\theta_b})$ and $y = a(\xi \sin{\theta_b} + \eta \cos{\theta_b})$.
The 50-orbit averages of the rotated surface density and the torque density are shown in the middle and right panels of Figure \ref{fig:density_fields} for runs with Mach 10 and 30. We find that in the Mach 30 case, the disk develops higher average surface density around the cavity wall and in the inner regions of the outer disk, and that this corresponds to a larger torque magnitude for $r > a$ (dotted lines in the right-most panel at $r = a$ and $r = r_{\rm cav} = 2.5 a$ delineate the inner, cavity, and outer-disk regions). The enhanced surface density in the cavity wall and inner portions of the outer disk can be seen clearly in the time-averaged radial surface density profiles shown in Figure \ref{fig:sig_profile}.

Aside from the change in surface density, it appears that both values of $\mathcal{M}$ yield essentially identical cavity morphology: a rotated ellipse whose near edges lie in the negative-torque quadrants, and whose far edges lie in the positive-torque quadrants\footnote{We note that this characteristic misaligned elliptic shape was pointed out by \citet[][see their Figure 1]{Dorazio2013} by considering massless test particles in the restricted 3-body problem.}.
This elongated and rotated density distribution tends to produce a net-negative gravitational torque due to the closer proximity of the gas in the negative-torque quadrants (upper left and lower right).

The growth of negative gravitational torques with increasing $\mathcal{M}$ could thus result from either of two effects: (1) the the pile-up of material around the cavity wall, or (2) the degree and orientation of the eccentric cavity. To quantify the cavity morphology, we have plotted iso-density contours of the surface density. The contours are shown in the left-most panels of Figure \ref{fig:density_fields}. The solid black contour corresponds to the minimum value on the color bar, and defines the shape of the cavity. We perform a least-squares fit of these contours to an ellipse, and then compare the best-fit rotation angle $\theta$ and eccentricity $\epsilon$ of each. The best-fit ellipses are displayed as the dotted-white curves in the middle-panels of Figure \ref{fig:density_fields}, and the fitting parameters for $\mathcal{M} = \{10, 30\}$ are shown in the top right corners of their respective rows. We find that for all Mach numbers considered in this study the best-fit ellipses for the time-averaged cavities have $\theta=0.34$ radians (from the vertical) and $\epsilon = [0.67, 0.69]$. 

We can thus rule out case (2) above; the cavity morphology and orientation are insensitive to Mach number, and the negative gravitational torque must be due to a greater mass of the negative-torque structure with Mach number.

To quantify the effect of gas accumulation around the cavity wall, we have calculated the angular distribution of the gravitational torque in the \emph{cavity region only}. We wish to compare this measurement with what is expected for an axisymmetric mass distribution. In a ring of uniform density material at radius $r$, the gas parcel at angle $\phi$ from the binary separation vector exerts a net torque per area
\begin{align}
    \frac{d T_g}{dA}(\phi) = A_\mathcal{M}\sin{\phi}\,\mathcal{D}(\phi; r) \, ,
 \label{eq:ang_torq}
\end{align}
where the amplitude $A_\mathcal{M}$ is determined by the surface density of the annulus, and
\begin{align}
	\begin{split}
		\mathcal{D}(\phi; r) &= r\, \bigg( \frac{1}{r_2^3} - \frac{1}{r_1^3} \bigg)\\ 
		&= \frac{r}{(r^2 + (a/2)^2 + ra\cos{\phi})^{3/2}} \\
		&- \frac{r}{(r^2 + (a/2)^2 - ra\cos{\phi})^{3/2}} \ .
	\end{split}
\end{align}
Integrated over $\phi$, Equation \ref{eq:ang_torq} would yield zero net torque on the binary, so in order to determine the amount by which our measured torque deviates from the axisymmetric prediction, we fit each measured angular torque distribution to Equation \ref{eq:ang_torq}. In our fits, we evaluate $\mathcal{D}(\phi;r)$ at $r = \bar r = 1.5a$, near the middle of the annulus defining the cavity region, and we confirm that the fit is not significantly effected by the choice for $\bar r$. Thus the sole fitting parameter is $A_\mathcal{M}$, the wave amplitude at Mach number $\mathcal{M}$.

\begin{figure}
    \includegraphics[width=\columnwidth]{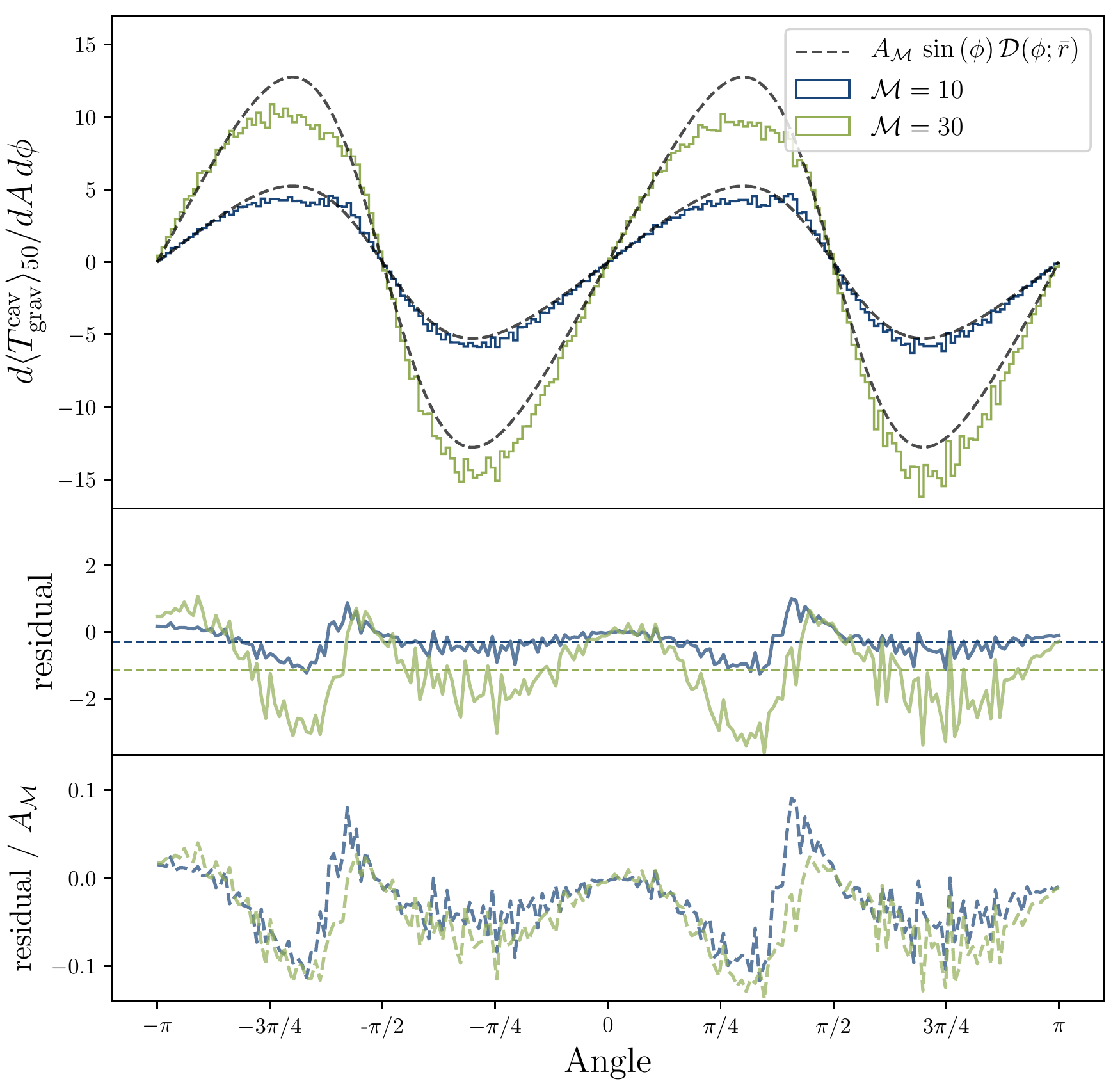}
    \protect\caption{The top panel shows angular torque distributions along a ring of material in the \emph{cavity region} ($a < r < 2.5\,a$)
    from the right panels of Figure \ref{fig:density_fields} (the 0-angle corresponds to the positive x-axis with respect to the origin, and positive angles coincide with counter-clockwise rotations from this ray). Each profile is fit to Equation \ref{eq:ang_torq}. The only free parameter is $A_\mathcal{M}$, the amplitude of a constant-density wave at that Mach number, which would yield zero net torque on the binary.  The two lower panels show, respectively, the residual between the measured torques and this fitting function and the residual as a percentage of the fit amplitude.  We see that the latter is nearly the same in both cases.  Therefore, the negative torques from the cavity region are the result of the cavity's characteristic geometry, and the larger torque magnitude at higher $\mathcal{M}$ simply corresponds to the higher overall surface-densities in this cavity structure.  In the case of $\mathcal{M} = 30$, these negative torques from the cavity overwhelm the positive torques from the inner region, causing the binary to inspiral.}
    \label{fig:ang_torq}
\end{figure}

The data and best-fit waves are shown in the top panel of Figure \ref{fig:ang_torq}. The middle panel shows the difference between the measured angular torque distribution and the axisymmetric prediction with best-fit $A_\mathcal{M}$. The horizontal lines denote the net residual, and the bottom panel shows this residual as a percentage of the best-fit amplitude. We see that the residual in the $\mathcal{M} = 30$ case is more negative than that of the $\mathcal{M} = 10$ case, consistent with the results shown in Figure \ref{fig:segmented}. However, as a percentage of the torque amplitude, the residuals are nearly identical. This further supports our finding that the cavity morphology and orientation are insensitive to Mach number, and that there is simply more gas packed into this structure at higher Mach numbers.

\subsection{Reason for the enhanced cavity density at high Mach number}
Now that we have established that the net torque is controlled by the quantity of mass concentrated around the cavity wall, we offer a possible explanation for this effect, and means by which it may be tested in the future. We propose that the radial mass distribution shown in Figure \ref{fig:sig_profile} is approaching a low-temperature limit, in which the cavity wall grows arbitrarily steep, and the surface density is sharply peaked at $r \approx 2.5a$. This low-temperature density structure has larger total mass than the smoother,
higher-temperature one, leading to maximal negative gravitational torque from the cavity wall. The reasons for this behavior in our simulations are still uncertain. Here we speculate on three possible contributing factors, leaving a rigorous diagnosis to a follow-up study.

First, the warmer disks exert a larger outward force due to the thermal pressure gradient. Adopting the analogy of a steady-state ``atmosphere'', the gas would need to have a shallower density profile to support itself against the ``weight'' (ram pressure) of the inward-drifting outer disk. The warmer disks with shallower density profile have their mass distributed farther from the binary, resulting in a smaller  negative torque from the cavity wall structure.

A second possibility is that the radial density profile is influenced by outward-propagating pressure waves. Any radial momentum deposited by the absorption of these waves tends to disperse gas outwards. If this picture is accurate, the length scale over which the radial $\Sigma$ profile is smoothed around the cavity wall should be connected to the attenuation length of the acoustic waves. These waves aught to shock and thermalize faster in colder disks, and indeed visual inspection of Figure \ref{fig:reliefs} suggests they propagate farthest in the warmer models (e.g. Mach 10, upper left).

Finally, a similar effect could be produced by the rejected gas streams themselves, rather than the cavity-wall medium into which they slam. We have found that a significant fraction of the gas approaching near the minidisks experiences a ``slingshot'' and is flung back towards the cavity wall. In the high Mach number runs, these rejected streams are denser and appear narrower, and carry higher specific angular momentum than the inward-traveling gas streams. The larger amount of mass, momentum, and angular momentum carried by the rejected gas may be caused by the properties of the shocks in the innermost regions, and can lead to the higher density peak seen in Figure \ref{fig:sig_profile} for the cold disks.

% ======================================================================
% ======================================================================
\section{Conclusions}
\label{sec:conclusions}

We have performed a suite of 2D isothermal hydrodynamics simulations of thin disk accretion onto an equal mass, circular binary using fixed mesh refinement in \texttt{Mara3}.  Historically, nearly all work on this topic has been performed at Mach number $\mathcal{M} = 10$ (or equivalently, constant scale-height $h/r = 0.1$) because this value has been computationally tractable and because it has become a useful benchmark for comparison across studies. However, disks observed around AGN are estimated to have Mach numbers of the order $\mathcal{M} \sim 10^2 - 10^3$. We explored the effect of raising the disk Mach number on the binary accretion rate, the angular momentum transfer between the disk and the binary, and the evolution of the binary separation. The conclusions of this work can be summarized as follows:

(1) We reproduce the result that binary accretion from finite disks, while not in a steady state, still manifest constant angular momentum transferred per unit accreted mass, $\ell_0$ (Figure ~\ref{fig:dots}). We do not consider an infinite disk in steady-state in this work, but for our fiducial run ($\mathcal{M} = 10, \alpha_{\rm eff}=0.1$ at $r = 2\, a$) 
we obtain an accretion eigenvalue ($\ell_0 = 0.81\,\sqrt{GMa}$) and migration rate per unit accreted mass ($da/dM = 2.17\,a\,M^{-1}$), in close agreement with the corresponding values presented in \citet[][see their Figure~9]{Munoz19}.

(2) The behavior of binaries accreting from a finite isothermal disk held at constant viscosity undergoes a transition from gas-driven outspiral to gas-mediated inspiral as their Mach number is increased (or disk thickness and temperature are decreased) from the commonly adopted value of $\mathcal{M} = 10$. For isothermal disks at constant viscosity, this transition occurs at $\mathcal{M} \sim 25$ (see Figures \ref{fig:mach_study}~and~\ref{fig:dots}). If the approximately linear decrease in the accretion eigenvalue $\ell_0$, the amount of angular momentum gained per unit accreted mass, holds out to $\mathcal{M} \sim 10^2 - 10^3$, this would imply that super-massive black hole binaries accreting from thin disks would experience strong negative torques and would be driven rapidly towards merger by the circumbinary disk.

(3) The angular momentum imparted to the binary per unit accreted mass, and consequently the change in binary separation due to gas accretion does not depend on Mach number (see Figure \ref{fig:delta_a}). Therefore, the transition from positive torques and expanding binaries to negative torques and shrinking binaries is due to the decrease in gravitational torques with increasing $\mathcal{M}$. By dissecting these gravitational torques into three regions, described in \S \ref{sec:torque}, we reproduced previous results that at $\mathcal{M} = 10$ the positive torques from the inner region are approximately double the magnitude of the net negative torques from the cavity and outer-disk regions. As we increase $\mathcal{M}$, while the growth of positive and negative torques in the inner and outer regions approximately offset, the negative torques in the cavity region grow by a much larger factor of nearly~4 (Figure \ref{fig:segmented}). By examining the orbit-averaged surface and torque density profiles, as well as the angular distribution of the gravitational torques in the cavity region in particular, we demonstrated that the torques from the cavity region are always negative because of the characteristic misaligned elongated shape of the time-averaged cavity. Furthermore, as $\mathcal{M}$ increases, more material builds up in the cavity walls and streamers. This overall increase in surface density leads to increasingly negative gravitational torques that eventually overwhelm positive torques from the gas in and near the minidisks, and results in the disk extracting angular momentum from the binary (see Figures \ref{fig:reliefs},~\ref{fig:density_fields},~and~\ref{fig:ang_torq}).

(4) We find that increasing $\mathcal{M}$ leads to a modest suppression in the binary accretion rate (Figure \ref{fig:disk_mass}).  This is in qualitative agreement with the results from \citet{Ragusa+2016} performed at constant turbulent viscosity parameter $\alpha$.
Our study, however, is performed at constant kinematic viscosity $\nu$, suggesting that the suppression is due to a decrease in pressure gradients ($\nabla P \sim \mathcal{M}^{-2}$) with increasing $\mathcal{M}$; as opposed to changes in viscosity.  This accretion suppression could imply that accreting super-massive binaries may be somewhat less luminous than their single-BH, AGN counterparts of comparable mass.

A principle finding of this work, following from conclusions (2) and (3), is that disks with $\mathcal{M} = 10$ are not truly ``thin-disks''.  Therefore, studies that use an artificially low value of the Mach number (or high $h/r$) will under-predict the density of gas in the cavity wall and streamers, and will thus underestimate the magnitude of the negative gravitational torques for $r > a$. These torques are crucial and mark the difference between expanding or ``stalled'' binaries and binaries that are driven to inspiral.

\subsection{Limitations and future work}
One major restriction in this work was that we assumed a constant value for the kinematic viscosity and allowed $\alpha$ to vary.  We briefly explored the effect of varying Mach number at constant $\alpha$ and found that it had little effect on our conclusions (see Figure \ref{fig:mach_study}). In future work we intend to explore this issue more thoroughly.

Another possibly important limitation of this paper is that we have only considered equal-mass binaries with zero orbital eccentricity. 
Varying the mass ratio and eccentricity can alter the cavity structure, and thus, may modify the torque balance.
A natural extension will be to consider binary orbits of varying mass ratio and eccentricity. 
Moreover, we could relax the assumption of small-amplitude orbital perturbations, and instead allow the binary to evolve self-consistently in response to gravitational and accretion forces. 

We have also assumed a locally isothermal equation of state, such that the CBD and minidisks are all held at the same temperature. This constraint ought to be relaxed with a more rigorous treatment of the thermodynamics, allowing the minidisks to exist out of thermal equilibrium with the CBD \citep{Farris15,Yike18} and accounting for physically realistic radiative cooling, as well as the dynamical effects of radiation pressure and radiative heating.
Specifically, when the streamers collide with the cavity edge they will shock and heat the gas, lowering the Mach number \citep{Yike18}. Proper treatment of shock heating and radiative cooling would allow the CBD to relax self-consistently to some Mach number and associated torque balance.

Lastly, while we expect two-dimensions to be sufficient for simulating thin-disks aligned with the binary orbital plane, the vertical structure could be important for disks of finite thickness and disk inclination could also change this picture; so 3D studies at different Mach numbers should eventually be carried out.

\vfill\null
\subsection{Implications for SMBHB evolution and observation}

As noted, recent works have suggested that SMBHBs embedded in thin accretion disks are driven away from merger, and may therefore stall. 
If this stalling occurs at large separations, well before the binary's orbital frequency enters the range of PTAs or  LISA,  it would
imply a relative dearth of compact binaries in the GW-emitting inspiral phase, possibly reducing the expected gravitational wave background (GWB), as well as the SMBHB merger rates detectable by LISA. If the trends with increasing Mach number (decreasing disk temperature and thickness) presented in this work hold, however, it would no longer suggest that such ``stalling'' is likely. Instead, the expectation is that gas-driven inspiral would produce a  population of compact massive binaries in the GW-emitting regime.

Only a handful of quasars have been identified with significant optical periodicities on a timescale of a $\sim$year, and put forward as massive black hole binary candidates. The presence of a gas--mediated rapid inspiral phase in SMBHB evolution could possibly also explain a relative paucity of such massive binaries with separations of order $\sim {\rm few}\, 10^{-2}{\rm pc}$.
If the gas torques are strong enough, the residence time in this gas--mediated phase could be sufficiently short so as to make observation of such a compact binary unlikely. However, the relative scarcity of these objects could also be explained by the suppression of the accretion rate at high Mach number, because, as noted here and also by \cite{Ragusa+2016}, this suppression could imply that accreting super-massive binaries are less luminous than their single-BH AGN counterparts of comparable mass (though shocks driven by the binary can produce additional luminosity components which could offset this dimming; \citealt{Farris15,Yike18}).

\section*{Acknowledgements}
We thank Daniel D'Orazio and Paul Duffell for helpful discussions and Mulin Ding for administering the Ria computing cluster at NYU.  Resources supporting this work were provided by the NASA High-End Computing (HEC) Program through the NASA Advanced Supercomputing (NAS) Division at Ames Research Center. We acknowledge support from NASA through ADAP grant 80NSSC18K1093 (to ZH) and {\it Swift} grant 80NSSC19K0149 (to ZH), and from the National Science Foundation through AST grant 1715661 (to AM and ZH).


\bibliography{paper}

\begin{thebibliography}{}
\makeatletter
\relax
\def\mn@urlcharsother{\let\do\@makeother \do\$\do\&\do\#\do\^\do\_\do\%\do\~}
\def\mn@doi{\begingroup\mn@urlcharsother \@ifnextchar [ {\mn@doi@}
  {\mn@doi@[]}}
\def\mn@doi@[#1]#2{\def\@tempa{#1}\ifx\@tempa\@empty \href
  {http://dx.doi.org/#2} {doi:#2}\else \href {http://dx.doi.org/#2} {#1}\fi
  \endgroup}
\def\mn@eprint#1#2{\mn@eprint@#1:#2::\@nil}
\def\mn@eprint@arXiv#1{\href {http://arxiv.org/abs/#1} {{\tt arXiv:#1}}}
\def\mn@eprint@dblp#1{\href {http://dblp.uni-trier.de/rec/bibtex/#1.xml}
  {dblp:#1}}
\def\mn@eprint@#1:#2:#3:#4\@nil{\def\@tempa {#1}\def\@tempb {#2}\def\@tempc
  {#3}\ifx \@tempc \@empty \let \@tempc \@tempb \let \@tempb \@tempa \fi \ifx
  \@tempb \@empty \def\@tempb {arXiv}\fi \@ifundefined
  {mn@eprint@\@tempb}{\@tempb:\@tempc}{\expandafter \expandafter \csname
  mn@eprint@\@tempb\endcsname \expandafter{\@tempc}}}

\bibitem[\protect\citeauthoryear{{Armitage} \& {Natarajan}}{{Armitage} \&
  {Natarajan}}{2002}]{Armitage2002}
{Armitage} P.~J.,  {Natarajan} P.,  2002, \mn@doi [\apjl] {10.1086/339770},
  \href {http://adsabs.harvard.edu/abs/2002ApJ...567L...9A} {567, L9}

\bibitem[\protect\citeauthoryear{{Armitage} \& {Natarajan}}{{Armitage} \&
  {Natarajan}}{2005}]{Armitage2005}
{Armitage} P.~J.,  {Natarajan} P.,  2005, \mn@doi [\apj] {10.1086/497108},
  \href {https://ui.adsabs.harvard.edu/abs/2005ApJ...634..921A} {634, 921}

\bibitem[\protect\citeauthoryear{{Barnes} \& {Hernquist}}{{Barnes} \&
  {Hernquist}}{1996}]{BarnesHernquist1996}
{Barnes} J.~E.,  {Hernquist} L.,  1996, \mn@doi [\apj] {10.1086/177957}, \href
  {http://adsabs.harvard.edu/abs/1996ApJ...471..115B} {471, 115}

\bibitem[\protect\citeauthoryear{{Begelman}, {Blandford}  \& {Rees}}{{Begelman}
  et~al.}{1980}]{Begel:Blan:Rees:1980}
{Begelman} M.~C.,  {Blandford} R.~D.,   {Rees} M.~J.,  1980, \mn@doi [\nat]
  {10.1038/287307a0}, 287, 307

\bibitem[\protect\citeauthoryear{{Charisi}, {Bartos}, {Haiman}, {Price-Whelan},
  {Graham}, {Bellm}, {Laher}  \& {M{\'a}rka}}{{Charisi}
  et~al.}{2016}]{Charisi+2016}
{Charisi} M.,  {Bartos} I.,  {Haiman} Z.,  {Price-Whelan} A.~M.,  {Graham}
  M.~J.,  {Bellm} E.~C.,  {Laher} R.~R.,   {M{\'a}rka} S.,  2016, \mn@doi
  [\mnras] {10.1093/mnras/stw1838}, \href
  {http://adsabs.harvard.edu/abs/2016MNRAS.463.2145C} {463, 2145}

\bibitem[\protect\citeauthoryear{{Cuadra}, {Armitage}, {Alexander}  \&
  {Begelman}}{{Cuadra} et~al.}{2009}]{Cuadra2009}
{Cuadra} J.,  {Armitage} P.~J.,  {Alexander} R.~D.,   {Begelman} M.~C.,  2009,
  \mn@doi [\mnras] {10.1111/j.1365-2966.2008.14147.x}, \href
  {http://adsabs.harvard.edu/abs/2009MNRAS.393.1423C} {393, 1423}

\bibitem[\protect\citeauthoryear{{D'Orazio}, {Haiman}  \&
  {MacFadyen}}{{D'Orazio} et~al.}{2013}]{Dorazio2013}
{D'Orazio} D.~J.,  {Haiman} Z.,   {MacFadyen} A.,  2013, \mn@doi [\mnras]
  {10.1093/mnras/stt1787}, \href
  {http://adsabs.harvard.edu/abs/2013MNRAS.436.2997D} {436, 2997}

\bibitem[\protect\citeauthoryear{{D'Orazio}, {Haiman}, {Duffell}, {MacFadyen}
  \& {Farris}}{{D'Orazio} et~al.}{2016}]{Dorazio2016}
{D'Orazio} D.~J.,  {Haiman} Z.,  {Duffell} P.,  {MacFadyen} A.,   {Farris} B.,
  2016, \mn@doi [\mnras] {10.1093/mnras/stw792}, \href
  {https://ui.adsabs.harvard.edu/\#abs/2016MNRAS.459.2379D} {459, 2379}

\bibitem[\protect\citeauthoryear{{Dotti}, {Colpi}, {Haardt}  \&
  {Mayer}}{{Dotti} et~al.}{2007}]{Dotti2007}
{Dotti} M.,  {Colpi} M.,  {Haardt} F.,   {Mayer} L.,  2007, \mn@doi [\mnras]
  {10.1111/j.1365-2966.2007.12010.x}, \href
  {http://adsabs.harvard.edu/abs/2007MNRAS.379..956D} {379, 956}

\bibitem[\protect\citeauthoryear{{Dotti}, {Ruszkowski}, {Paredi}, {Colpi},
  {Volonteri}  \& {Haardt}}{{Dotti} et~al.}{2009}]{Dotti2009}
{Dotti} M.,  {Ruszkowski} M.,  {Paredi} L.,  {Colpi} M.,  {Volonteri} M.,
  {Haardt} F.,  2009, \mn@doi [\mnras] {10.1111/j.1365-2966.2009.14840.x},
  \href {http://adsabs.harvard.edu/abs/2009MNRAS.396.1640D} {396, 1640}

\bibitem[\protect\citeauthoryear{{Dressler} \& {Richstone}}{{Dressler} \&
  {Richstone}}{1988}]{Dressler1988}
{Dressler} A.,  {Richstone} D.~O.,  1988, \mn@doi [\apj] {10.1086/165930},
  \href {http://adsabs.harvard.edu/abs/1988ApJ...324..701D} {324, 701}

\bibitem[\protect\citeauthoryear{Escala, Larson, Coppi  \& Mardones}{Escala
  et~al.}{2005}]{Escala2005}
Escala A.,  Larson R.~B.,  Coppi P.~S.,   Mardones D.,  2005, \mn@doi [The
  Astrophysical Journal] {10.1086/431747}, 630, 152

\bibitem[\protect\citeauthoryear{Farris, Duffell, MacFadyen  \& Haiman}{Farris
  et~al.}{2015}]{Farris15}
Farris B.~D.,  Duffell P.,  MacFadyen A.~I.,   Haiman Z.,  2015, \mn@doi
  [Monthly Notices of the Royal Astronomical Society: Letters]
  {10.1093/mnrasl/slu184}, 447, L80

\bibitem[\protect\citeauthoryear{{Ferrarese} \& {Ford}}{{Ferrarese} \&
  {Ford}}{2005}]{Ferrarese2005}
{Ferrarese} L.,  {Ford} H.,  2005, \mn@doi [\ssr] {10.1007/s11214-005-3947-6},
  \href {http://adsabs.harvard.edu/abs/2005SSRv..116..523F} {116, 523}

\bibitem[\protect\citeauthoryear{{Fiacconi}, {Mayer}, {Ro{\v s}kar}  \&
  {Colpi}}{{Fiacconi} et~al.}{2013}]{Fiacconi+2013}
{Fiacconi} D.,  {Mayer} L.,  {Ro{\v s}kar} R.,   {Colpi} M.,  2013, \mn@doi
  [\apjl] {10.1088/2041-8205/777/1/L14}, \href
  {http://adsabs.harvard.edu/abs/2013ApJ...777L..14F} {777, L14}

\bibitem[\protect\citeauthoryear{{Gould} \& {Rix}}{{Gould} \&
  {Rix}}{2000}]{Gould2000}
{Gould} A.,  {Rix} H.-W.,  2000, \mn@doi [\apjl] {10.1086/312562}, \href
  {https://ui.adsabs.harvard.edu/abs/2000ApJ...532L..29G} {532, L29}

\bibitem[\protect\citeauthoryear{{Graham} et~al.,}{{Graham}
  et~al.}{2015}]{Graham+2015}
{Graham} M.~J.,  et~al., 2015, \mn@doi [\mnras] {10.1093/mnras/stv1726}, \href
  {http://adsabs.harvard.edu/abs/2015MNRAS.453.1562G} {453, 1562}

\bibitem[\protect\citeauthoryear{{Haiman}, {Kocsis}  \& {Menou}}{{Haiman}
  et~al.}{2009}]{HKM09}
{Haiman} Z.,  {Kocsis} B.,   {Menou} K.,  2009, \mn@doi [\apj]
  {10.1088/0004-637X/700/2/1952}, \href
  {http://adsabs.harvard.edu/abs/2009ApJ...700.1952H} {700, 1952}

\bibitem[\protect\citeauthoryear{{Hubeny}, {Blaes}, {Krolik}  \&
  {Agol}}{{Hubeny} et~al.}{2001}]{Hubeny2001}
{Hubeny} I.,  {Blaes} O.,  {Krolik} J.~H.,   {Agol} E.,  2001, \mn@doi [\apj]
  {10.1086/322344}, \href
  {https://ui.adsabs.harvard.edu/abs/2001ApJ...559..680H} {559, 680}

\bibitem[\protect\citeauthoryear{{Kelley}, {Blecha}  \& {Hernquist}}{{Kelley}
  et~al.}{2017a}]{Kelley+2017a}
{Kelley} L.~Z.,  {Blecha} L.,   {Hernquist} L.,  2017a, \mn@doi [\mnras]
  {10.1093/mnras/stw2452}, \href
  {https://ui.adsabs.harvard.edu/abs/2017MNRAS.464.3131K} {464, 3131}

\bibitem[\protect\citeauthoryear{{Kelley}, {Blecha}, {Hernquist}, {Sesana}  \&
  {Taylor}}{{Kelley} et~al.}{2017b}]{Kelley+2017b}
{Kelley} L.~Z.,  {Blecha} L.,  {Hernquist} L.,  {Sesana} A.,   {Taylor} S.~R.,
  2017b, \mn@doi [\mnras] {10.1093/mnras/stx1638}, \href
  {https://ui.adsabs.harvard.edu/abs/2017MNRAS.471.4508K} {471, 4508}

\bibitem[\protect\citeauthoryear{{Kelley}, {Haiman}, {Sesana}  \&
  {Hernquist}}{{Kelley} et~al.}{2019}]{Kelley+2019}
{Kelley} L.~Z.,  {Haiman} Z.,  {Sesana} A.,   {Hernquist} L.,  2019, \mn@doi
  [\mnras] {10.1093/mnras/stz150}, \href
  {https://ui.adsabs.harvard.edu/abs/2019MNRAS.485.1579K} {485, 1579}

\bibitem[\protect\citeauthoryear{{Kocsis} \& {Sesana}}{{Kocsis} \&
  {Sesana}}{2011}]{KocsisSesana2011}
{Kocsis} B.,  {Sesana} A.,  2011, \mn@doi [\mnras]
  {10.1111/j.1365-2966.2010.17782.x}, \href
  {https://ui.adsabs.harvard.edu/abs/2011MNRAS.411.1467K} {411, 1467}

\bibitem[\protect\citeauthoryear{{Kocsis}, {Haiman}  \& {Loeb}}{{Kocsis}
  et~al.}{2012a}]{Kocsis+2012a}
{Kocsis} B.,  {Haiman} Z.,   {Loeb} A.,  2012a, \mn@doi [\mnras]
  {10.1111/j.1365-2966.2012.22129.x}, 427, 2660

\bibitem[\protect\citeauthoryear{{Kocsis}, {Haiman}  \& {Loeb}}{{Kocsis}
  et~al.}{2012b}]{Kocsis+2012b}
{Kocsis} B.,  {Haiman} Z.,   {Loeb} A.,  2012b, \mn@doi [\mnras]
  {10.1111/j.1365-2966.2012.22118.x}, 427, 2680

\bibitem[\protect\citeauthoryear{{Kormendy} \& {Ho}}{{Kormendy} \&
  {Ho}}{2013}]{KormendyHo2013}
{Kormendy} J.,  {Ho} L.~C.,  2013, \mn@doi [\araa]
  {10.1146/annurev-astro-082708-101811}, \href
  {http://adsabs.harvard.edu/abs/2013ARA%26A..51..511K} {51, 511}

\bibitem[\protect\citeauthoryear{{Kormendy} \& {Richstone}}{{Kormendy} \&
  {Richstone}}{1995}]{Kormendy1995}
{Kormendy} J.,  {Richstone} D.,  1995, \mn@doi [\araa]
  {10.1146/annurev.aa.33.090195.003053}, \href
  {http://adsabs.harvard.edu/abs/1995ARA%26A..33..581K} {33, 581}

\bibitem[\protect\citeauthoryear{{Krolik}}{{Krolik}}{1999}]{krolikbook}
{Krolik} J.~H.,  1999, {Active galactic nuclei: from the central black hole to
  the galactic environment}.
Princeton University Press, Princeton, New Jersey

\bibitem[\protect\citeauthoryear{{MacFadyen} \&
  {Milosavljevi{\'c}}}{{MacFadyen} \&
  {Milosavljevi{\'c}}}{2008}]{2008ApJ...672...83M}
{MacFadyen} A.~I.,  {Milosavljevi{\'c}} M.,  2008, \mn@doi [\apj]
  {10.1086/523869}, \href {http://adsabs.harvard.edu/abs/2008ApJ...672...83M}
  {672, 83}

\bibitem[\protect\citeauthoryear{{Mayer}, {Kazantzidis}, {Madau}, {Colpi},
  {Quinn}  \& {Wadsley}}{{Mayer} et~al.}{2007}]{Mayer+2007}
{Mayer} L.,  {Kazantzidis} S.,  {Madau} P.,  {Colpi} M.,  {Quinn} T.,
  {Wadsley} J.,  2007, \mn@doi [Science] {10.1126/science.1141858}, \href
  {http://adsabs.harvard.edu/abs/2007Sci...316.1874M} {316, 1874}

\bibitem[\protect\citeauthoryear{{Milosavljevi{\'c}} \&
  {Merritt}}{{Milosavljevi{\'c}} \& {Merritt}}{2001}]{2001ApJ...563...34M}
{Milosavljevi{\'c}} M.,  {Merritt} D.,  2001, \mn@doi [\apj] {10.1086/323830},
  \href {http://adsabs.harvard.edu/abs/2001ApJ...563...34M} {563, 34}

\bibitem[\protect\citeauthoryear{{Milosavljevi{\'c}} \&
  {Merritt}}{{Milosavljevi{\'c}} \& {Merritt}}{2003}]{fpp2003}
{Milosavljevi{\'c}} M.,  {Merritt} D.,  2003, \mn@doi [\apj] {10.1086/378086},
  \href {https://ui.adsabs.harvard.edu/abs/2003ApJ...596..860M} {596, 860}

\bibitem[\protect\citeauthoryear{{Milosavljevi{\'c}} \&
  {Merritt}}{{Milosavljevi{\'c}} \& {Merritt}}{2005}]{Merritt2005}
{Milosavljevi{\'c}} M.,  {Merritt} D.,  2005, \mn@doi [Living Reviews in
  Relativity] {10.12942/lrr-2005-8}, \href
  {http://adsabs.harvard.edu/abs/2005LRR.....8....8M} {8}

\bibitem[\protect\citeauthoryear{{Miranda}, {Mu{\~n}oz}  \& {Lai}}{{Miranda}
  et~al.}{2017}]{MML17}
{Miranda} R.,  {Mu{\~n}oz} D.~J.,   {Lai} D.,  2017, \mn@doi [\mnras]
  {10.1093/mnras/stw3189}, \href
  {http://adsabs.harvard.edu/abs/2017MNRAS.466.1170M} {466, 1170}

\bibitem[\protect\citeauthoryear{Moody, Shi  \& Stone}{Moody
  et~al.}{2019}]{Moody19}
Moody M. S.~L.,  Shi J.-M.,   Stone J.~M.,  2019, \mn@doi [The Astrophysical
  Journal] {10.3847/1538-4357/ab09ee}, 875, 66

\bibitem[\protect\citeauthoryear{{Mu{\~n}oz}, {Miranda}  \& {Lai}}{{Mu{\~n}oz}
  et~al.}{2019}]{MML19}
{Mu{\~n}oz} D.~J.,  {Miranda} R.,   {Lai} D.,  2019, \mn@doi [\apj]
  {10.3847/1538-4357/aaf867}, \href
  {http://adsabs.harvard.edu/abs/2019ApJ...871...84M} {871, 84}

\bibitem[\protect\citeauthoryear{{Mu{\~n}oz}, {Lai}, {Kratter}  \& {Mirand
  a}}{{Mu{\~n}oz} et~al.}{2020}]{Munoz19}
{Mu{\~n}oz} D.~J.,  {Lai} D.,  {Kratter} K.,   {Mirand a} R.,  2020, \mn@doi
  [\apj] {10.3847/1538-4357/ab5d33}, \href
  {https://ui.adsabs.harvard.edu/abs/2020ApJ...889..114M} {889, 114}

\bibitem[\protect\citeauthoryear{{Rafikov}}{{Rafikov}}{2016}]{Rafikov2016}
{Rafikov} R.~R.,  2016, \mn@doi [\apj] {10.3847/0004-637X/827/2/111}, \href
  {http://adsabs.harvard.edu/abs/2016ApJ...827..111R} {827, 111}

\bibitem[\protect\citeauthoryear{{Ragusa}, {Lodato}  \& {Price}}{{Ragusa}
  et~al.}{2016}]{Ragusa+2016}
{Ragusa} E.,  {Lodato} G.,   {Price} D.~J.,  2016, \mn@doi [\mnras]
  {10.1093/mnras/stw1081}, \href
  {http://adsabs.harvard.edu/abs/2016MNRAS.460.1243R} {460, 1243}

\bibitem[\protect\citeauthoryear{{Roedig}, {Sesana}, {Dotti}, {Cuadra},
  {Amaro-Seoane}  \& {Haardt}}{{Roedig} et~al.}{2012}]{Roedig2012}
{Roedig} C.,  {Sesana} A.,  {Dotti} M.,  {Cuadra} J.,  {Amaro-Seoane} P.,
  {Haardt} F.,  2012, \mn@doi [\aap] {10.1051/0004-6361/201219986}, \href
  {http://adsabs.harvard.edu/abs/2012A%26A...545A.127R} {545, A127}

\bibitem[\protect\citeauthoryear{{Roos}}{{Roos}}{1981}]{Roos1981}
{Roos} N.,  1981, \aap, \href
  {http://adsabs.harvard.edu/abs/1981A%26A...104..218R} {104, 218}

\bibitem[\protect\citeauthoryear{{Shakura} \& {Sunyaev}}{{Shakura} \&
  {Sunyaev}}{1973}]{SS1973}
{Shakura} N.~I.,  {Sunyaev} R.~A.,  1973, \aap, \href
  {http://adsabs.harvard.edu/abs/1973A%26A....24..337S} {24, 337}

\bibitem[\protect\citeauthoryear{Shi, Krolik, Lubow  \& Hawley}{Shi
  et~al.}{2012}]{Shi+2012}
Shi J.-M.,  Krolik J.~H.,  Lubow S.~H.,   Hawley J.~F.,  2012, \apj, 749, 118

\bibitem[\protect\citeauthoryear{{Souza Lima}, {Mayer}, {Capelo}  \&
  {Bellovary}}{{Souza Lima} et~al.}{2017}]{SouzaLima+2017}
{Souza Lima} R.,  {Mayer} L.,  {Capelo} P.~R.,   {Bellovary} J.~M.,  2017,
  \mn@doi [\apj] {10.3847/1538-4357/aa5d19}, \href
  {http://adsabs.harvard.edu/abs/2017ApJ...838...13S} {838, 13}

\bibitem[\protect\citeauthoryear{{Springel}, {Di Matteo}  \&
  {Hernquist}}{{Springel} et~al.}{2005}]{Springel2005}
{Springel} V.,  {Di Matteo} T.,   {Hernquist} L.,  2005, \mn@doi [\apjl]
  {10.1086/428772}, \href {http://adsabs.harvard.edu/abs/2005ApJ...620L..79S}
  {620, L79}

\bibitem[\protect\citeauthoryear{{Syer} \& {Clarke}}{{Syer} \&
  {Clarke}}{1995}]{SyerClarke1995}
{Syer} D.,  {Clarke} C.~J.,  1995, \mn@doi [\mnras] {10.1093/mnras/277.3.758},
  \href {http://adsabs.harvard.edu/abs/1995MNRAS.277..758S} {277, 758}

\bibitem[\protect\citeauthoryear{{Tang}, {MacFadyen}  \& {Haiman}}{{Tang}
  et~al.}{2017}]{Yike17}
{Tang} Y.,  {MacFadyen} A.,   {Haiman} Z.,  2017, \mn@doi [\mnras]
  {10.1093/mnras/stx1130}, \href
  {http://adsabs.harvard.edu/abs/2017MNRAS.469.4258T} {469, 4258}

\bibitem[\protect\citeauthoryear{{Tang}, {Haiman}  \& {MacFadyen}}{{Tang}
  et~al.}{2018}]{Yike18}
{Tang} Y.,  {Haiman} Z.,   {MacFadyen} A.,  2018, \mn@doi [\mnras]
  {10.1093/mnras/sty423}, \href
  {http://adsabs.harvard.edu/abs/2018MNRAS.476.2249T} {476, 2249}

\bibitem[\protect\citeauthoryear{{Vasiliev}, {Antonini}  \&
  {Merritt}}{{Vasiliev} et~al.}{2015}]{Vasiliev+2015}
{Vasiliev} E.,  {Antonini} F.,   {Merritt} D.,  2015, \mn@doi [\apj]
  {10.1088/0004-637X/810/1/49}, \href
  {https://ui.adsabs.harvard.edu/abs/2015ApJ...810...49V} {810, 49}

\bibitem[\protect\citeauthoryear{{Woods} et~al.,}{{Woods}
  et~al.}{2019}]{Woods+2019}
{Woods} T.~E.,  et~al., 2019, \mn@doi [\pasa] {10.1017/pasa.2019.14}, \href
  {https://ui.adsabs.harvard.edu/abs/2019PASA...36...27W} {36, e027}

\bibitem[\protect\citeauthoryear{{Zrake} \& {MacFadyen}}{{Zrake} \&
  {MacFadyen}}{2012}]{Mara}
{Zrake} J.,  {MacFadyen} A.~I.,  2012, \mn@doi [\apj]
  {10.1088/0004-637X/744/1/32}, \href
  {https://ui.adsabs.harvard.edu/abs/2012ApJ...744...32Z} {744, 32}

\bibitem[\protect\citeauthoryear{Zrake, Tiede, MacFadyen  \& Haiman}{Zrake
  et~al.}{2021}]{Zrake2020}
Zrake J.,  Tiede C.,  MacFadyen A.,   Haiman Z.,  2021, \mn@doi [The
  Astrophysical Journal Letters] {10.3847/2041-8213/abdd1c}, 909, L13

\makeatother
\end{thebibliography}
\bibliographystyle{mnras}
\label{lastpage}



\end{document}